\documentclass{emulateapj} 
\journalinfo{The Astrophysical Journal, 684:691--706, 2008 September 1}
\submitted{Received 2007 December 21; accepted 2008 May 10}
\newcommand{\myemail}{josan@mso.anu.edu.au}
\newcommand{\mr}{\mathrm}
\usepackage{lscape}
\shorttitle{Comparison of the Sun to other Stars}
\shortauthors{Robles et al.}
\begin{document}
\title{A comprehensive comparison of the Sun to other stars: searching for self-selection effects}
%%%%%%%%%%%%%%%%%%%%%%%%%%%%%%%%%%%%%%%%%%%%%%%
%Authors & Institutions 
\author{Jos\'e A. Robles\altaffilmark{1}, Charles H. Lineweaver\altaffilmark{1}, Daniel Grether\altaffilmark{2}, 
Chris Flynn\altaffilmark{3}, Chas A. Egan\altaffilmark{2,4}, Michael B. Pracy\altaffilmark{4},
Johan Holmberg\altaffilmark{5} and Esko Gardner\altaffilmark{3}}

\altaffiltext{1}{Planetary Science Institute, Research School of Astronomy \& Astrophysics and Research School of Earth Sciences, The Australian National University, Canberra Australia; \myemail.}
\altaffiltext{2}{University of New South Wales, Sydney, Australia.}
\altaffiltext{3}{Tuorla Observatory, University of Turku, Finland.}
\altaffiltext{4}{Research School of Astronomy \& Astrophysics,  The Australian National University, Canberra, Australia.}
\altaffiltext{5}{Max Planck Institute for Astronomy, Heidelberg, Germany.}
%%%%%%%%%%%%%%%%%%%%%%%%%%%%%%%%%%%%%%%%%%%%%%%
%%%%%%%%%%%%%%%%%%%%%%%%%%%%%%%%%%%%%%%%%%%%%%%
% ABSTRACT
\begin{abstract}
If the origin of life and the evolution of observers on a planet is favoured by atypical properties
of a planet's host star, we would expect our Sun to be atypical with respect to such properties. 
The Sun has been described by previous studies as both typical and atypical. In an effort to reduce 
this ambiguity and quantify how typical the Sun is, we identify 11 maximally-independent 
properties that have plausible correlations with habitability, and that have been observed by, or 
can be derived from, sufficiently large, currently available and representative stellar surveys.
By comparing solar values for the 11 properties, to the resultant stellar distributions,
we make the most comprehensive comparison of the Sun to other stars. The two most atypical properties 
of the Sun are its mass and orbit. The Sun is more massive than $95 \pm 2\%$ of nearby stars and 
its orbit around the Galaxy is less eccentric than $93 \pm 1 \%$ of FGK stars within $40$ parsecs. 
Despite these apparently atypical properties, a $\chi^{2}$-analysis of the Sun's values for 
11 properties, taken together, yields a solar $\chi^2_{\odot} = 8.39 \pm 0.96$.  
If a star is chosen at random,  the probability that it will have a lower value ($\sim$ be more typical) 
than the Sun, with respect to the 11 properties analyzed here,    
is only  $29 \pm 11 \%$.
These values quantify, and are consistent with, the idea that the Sun is a typical star.
If we have sampled all reasonable properties associated with habitability,
our result suggests that there are no special requirements for a star to 
host a planet with life.

\end{abstract}

%%%%%%%%%%%%%%%%%%%%%%%%%%%%%%%%%%%%%%%%%%%%%%%
% INTRODUCTION
\section{INTRODUCTION}

If the properties of the Sun are consistent with the idea that the Sun was randomly  
selected from all stars, this would indicate that life needs nothing special from its host star 
and would support the idea that life may be common in the universe.  More particularly, if 
there is nothing special about the Sun, we have little reason to limit our life-hunting efforts 
to planets orbiting Sun-like stars. As an example of the type of anthropic reasoning we are using, 
consider the following situation. Suppose uranium (a low abundance element in the Solar 
System and in the universe) was central to the biochemistry of life on Earth.  
Further, suppose that a comparison of our Sun to other stars showed that the 
Sun had more uranium than any other star.  How should we interpret this fact? 
The most reasonable way to proceed would be to try to evaluate the probability 
that such a coincidence happened by chance and to determine whether we are justified in 
reading some importance into it.  Although a correlation does not necessarily 
imply cause, we think that a correlation between the Sun's 
anomalous feature and life's fundamental chemistry would be giving us 
important clues about the conditions necessary for life. Specifically, the search 
for life around other stars as envisioned by the NASA's Terrestrial Planet 
Finder or ESA's Darwin Project and as currently underway with SETI would 
change the strategy to focus on the most uranium-rich stars.  
Another example: Suppose the Sun had the highest [Fe/H] of all the stars that 
had ever been observed.  Then high [Fe/H] would be strongly implicated as a 
precondition for our existence, possibly by playing a crucial role in terrestrial 
planet formation.  These are exaggerated examples of the more subtle 
correlations that a detailed and comprehensive comparison of the Sun with 
other stars could reveal.

Whether the Sun is a typical or atypical star with respect 
to 1 or a few properties has been addressed in previous studies. Using an approach similar 
to ours (comparing solar to stellar properties from particular samples), some studies have 
suggested that the Sun is a typical star \citep{Gustafsson98, Allendeprieto06}, while other studies 
have suggested that the Sun is an atypical star \citep{Gonzalez99a,Gonzalez99b,Gonzalez01}. 
This apparent disagreement arises from three problems:
\begin{description}
\item{1.} Language used to describe whether the Sun is, or is not typical, is often 
confusingly qualitative. For example, reporting the Sun as ``metal-rich'', can mean that the Sun 
is \textit{significantly} more metal-rich than other stars (e.g.\ more metal-rich than $80\%$ of 
other stars) or it can  mean that the Sun is \textit{insignificantly} metal-rich (e.g.\ more 
metal-rich than $51\%$ of other stars).

\item{2.} Selection effects: Stellar samples chosen for the comparison can be biased
with respect to the property of interest.  

\item{3.} Inclusion (or exclusion) of stellar properties for which it is
suspected or known that the Sun is atypical, will make the Sun appear more atypical (or typical). 
\end{description}

In this paper we address problem 1 by using only quantitative measures when comparing 
the Sun's properties to other stars. Our main interest is to move beyond the qualitative 
assessment of the Sun as either typical or atypical, and obtain a more precise quantification 
of the degree of the Sun's (a)typicality. In other words, we want to answer the question 
`How typical is the Sun?' rather than `Is the Sun typical or not?' There are at least
two ways to quantify how typical the Sun is. This can be done for individual parameters 
by determining how many stars have values below or above the solar value (Table \ref{table:sum}).  
This can also be done by a joint analysis of multiple parameters (Table \ref{table:chiimprovedp}).
If there are several subtle factors that have some influence over habitability, a quantitative 
joint analysis of the Sun's properties may allow us to identify these factors without invoking 
largely speculative arguments linking specific properties to habitability.

With respect to problem 2, most previous analyses have compared the Sun to subsets of 
Sun-like stars selected to be Sun-like with respect to 1 or more parameters. 
In such analyses, the Sun will appear typical with respect to any parameter(s) correlated with 1
of the pre-selected Sun-like parameters.
For example, elemental abundances [X/H] are correlated with 
metallicity\footnote{Metallicity: [Fe/H] is the fractional 
abundance of Fe relative to hydrogen, compared to the same ratio in the Sun: \mbox{[Fe/H] 
$\equiv \log(\mr{Fe/H})_{\star} - \log(\mr{Fe/H})_{\sun}$}} [Fe/H]. The sample of \cite{Edvardsson93} was selected to have a wide range of [Fe/H]. 
This produced a metallicity  distribution unrepresentative of stars in general. 
Recognizing this, \cite{Edvardsson93} conditioned on solar metallicity,
[Fe/H$] \approx 0$ and then 
compared solar abundances for 12 elements to the abundances in a group of
nearby stars with solar iron abundance, solar age and solar galactocentric radius. 
They found the Sun to be ``a quite typical star for its 
metallicity, age and galactic orbit''. 
Similarly, \cite{Gustafsson98}, after comparing various properties 
of the Sun to solar-type stars (stars of similar mass and age), concluded that the Sun seems 
very normal for its mass and age; ``The Sun, to a remarkable degree, is solar type''.
The stellar samples we use for comparison with the Sun are, in our judgement, the 
least-biased samples currently available for such a comparison.

To address problem 3, in Section \ref{sec:data}
we compare the Sun to other stars using a large number  (11) of maximally-independent 
properties with plausible correlations with habitability.  These properties can be observed or 
derived for a sufficiently large, representative stellar sample (Table \ref{table:samples1}).
Any property of the Sun or its environment which must be special to allow habitability would show up in our analysis.
However, in contrast to previous analyses which have looked for solar anomalies with respect to 
individual properties, we perform a joint analysis that enables us to quantify how typical 
the solar values are, taken as a group. In Section \ref{sec:jointa}, the differences between the 
solar values and the stellar samples' medians are used to perform first a simple and then an 
improved version of a $\chi^{2}$-analysis to estimate whether the solar values are characteristic 
of a star selected at random from the stellar samples. The results of our joint analysis 
are presented in Figure \ref{fig:sum} of Section \ref{sec:results}.  We find that the solar 
values, taken as a group, are consistent with the Sun being a random star. However, there are important
caveats to this interpretation associated with the compromise between the number of properties
analyzed, and their plausibility of being correlated with habitability. In  Sections \ref{sec:discuss} and 
\ref{sec:conclusions}  we discuss these caveats and summarize. We discuss the levels of correlation
between our 11 properties in Appendix A.

%%%%%%%%%% table data sets properties 
\thispagestyle{empty}
\setlength{\voffset}{30mm}
% {\rotate
\begin{deluxetable*}{lllcccrcrl}%[!ttt]  
  \tabletypesize{\footnotesize}
  \tablewidth{0pt}
  \tablecaption{Samples used to produce the stellar distributions plotted in Figures 1--10.}
  \tablehead{
    \colhead{Figure}  & \colhead{Property}  & \colhead{Range}  &  \colhead{Median}   & \colhead{$\sigma_{68}$\tablenotemark{\mbox{\dag}}}  & \colhead{Solar}  & \colhead{\# Stars} & \colhead{Spectral}   & \colhead{$d_{\mr{max}}$}  & \colhead{Source} \\
    &  &  & \colhead{$\mu_{1/2}$}   &   & \colhead{Value} &    & \colhead{Type}  &  \colhead{[pc]}    &          
  }
  \startdata
  \ref{fig:mass}     &Mass [M$_{\odot}$]              &$0.08$ -- $2$          & 0.33         &0.37       &      1                   & 125   &A1--M7  &   7.1  & \citealt{Henry06} (RECONS)     \\
  \ref{fig:age}      &Age  [Gyr]                     &$0$ -- $15$            & 5.4          &3.25        &  4.9$^{+3.1}_{-2.7}$\tablenotemark{a}    & 552   &F8--K2  &   200  & \citealt{Rochapinto00a}        \\
  \ref{fig:fe}       &[Fe/H]                         &$-1.20$ -- $+0.46$      & -0.08       &0.20        &    0                      & 453   &F7--K3  &   25   & \citealt{Grether07}            \\           
  \ref{fig:ratios}A  &[C/O]                          &$-0.22$ -- $+0.32$      & 0.07        &0.09        &    0                    & 256   &FG      &  150   & \tablenotemark{b}G99, R03, BF06             \\ 
  \ref{fig:ratios}B  &[Mg/Si]                        &$-0.18$ -- $+0.14$      & 0.01        &0.04        &    0                  & 231   &FG      &  150   & \tablenotemark{c}R03, B05                   \\
  \ref{fig:vsini}    &$v\sin i$ [\mbox{km s$^{-1}$}]  &$0$ -- $36$            & 2.51         &1.27       & $1.28$\tablenotemark{d}             & 276   &F8--K2 &   80   & \tablenotemark{e}\citealt{Valenti05}      \\        
  \ref{fig:e}        &$e$                            &$0$ --  $1$            & 0.10         &0.05        & $0.036 \pm 0.002$\tablenotemark{f}  & 1,987 &A5--K2  &   40   & \tablenotemark{g}\citealt{Nordstrom04}      \\ 
  \ref{fig:z}        &$Z_{\mr{max}}$ [kpc]             &$0$ --   $9.60$         & 0.14         &0.10     & $0.104 \pm 0.006$\tablenotemark{h} & 1,987 &A5--K2  &   40   & \tablenotemark{g}\citealt{Nordstrom04}      \\
  \ref{fig:radius}   &R$_{\mr{Gal}}$ [kpc]             &$0$ -- $30$            & 4.9          &5.03      &$7.62 \pm 0.32$\tablenotemark{i}    & ---   & ---    & 50,000 & \tablenotemark{j}BS80, G96, E05             \\    
  \ref{fig:galaxy}   &M$_{\mr{gal}}$ [M$_{\odot}$]\tablenotemark{k}   &$10^{7}$ -- $10^{12}$ &$10^{10.2}$     &0.47     &$10^{10.55 \pm 0.16 }$ & ---   & ---    &$10^{7}$& \tablenotemark{l}D94, CB99, L00, BJ01, J03  \\           
  \ref{fig:group}    &M$_{\mr{group}}$ [M$_{\odot}$]\tablenotemark{k} &$10^{9}$ -- $10^{13}$ &$10^{11.1}$      &0.47    &$10^{10.91 \pm 0.07}$  & ---   & ---    &$10^{7}$& \citealt{Ekev04}       \\           
  \enddata
  \tablenotetext{$\dagger$}{}{Characteristic width of distribution in the direction of the solar value.}
  \tablenotetext{a}{\cite{Wright04}, (see footnote in Sec.~\ref{sec:age}).}
  \tablenotetext{b}{G99: \citealt{Gustafsson99}, R03: \citealt{Reddy03}, BF06: \citealt{Bensby06}.}
  \tablenotetext{c}{R03: \citealt{Reddy03}, B05: \citealt{Bensby05}.}
  \tablenotetext{d}{Solar rotational velocity corrected for random inclination (see Sec.~\ref{sec:rotvel}).}
  \tablenotetext{e}{Sub-set of stars within the mass range: 0.9 M$_{\odot} \leq M \leq 1.1$ M$_{\odot}$.}
  \tablenotetext{f}{Calculated using the solar galactic motion \citep{dehnen98} and the Galactic potential (see Sec.~\ref{sec:orbital}).}
  \tablenotetext{g}{Sub-set of volume complete A5--K2 stars within 40 pc.}
  \tablenotetext{h}{Integrated solar orbit in the Galactic potential of \cite{flynn96} (see Sec.~\ref{sec:orbital}).}
  \tablenotetext{i}{\citealt{Eisenhauer05}.}
  \tablenotetext{j}{BS80: \citealt{Bahcall80}, G96: \citealt{Gould96}, E05: \citealt{Eisenhauer05}.}
  \tablenotetext{k}{Stellar mass, not total baryonic mass, nor total mass.}
  \tablenotetext{l}{D94: \citealt{driver94}, CB99: \citealt{courteau99}, L00: \citealt{lovedayj00}, BJ01: \citealt{bell01}, J03: \citealt{jarrett03}.}
  \label{table:samples1}
\end{deluxetable*}
% }
\setlength{\voffset}{0mm}
%%%%%%%%%%% 
\section{Stellar Samples and Solar Values} \label{sec:data}

We are looking for a signal associated with a prerequisite for, or a property that favors, 
the origin and evolution of life (see \citealt{Gustafsson98} for a brief discussion of this idea).
If we indiscriminately include many properties with little or 
no plausible correlation with habitability, we run the risk of diluting any  potential signal.  
If we choose only a few properties based on previous knowledge that the Sun is anomalous
with respect to those properties, we are making a useful quantification but we are 
unable to address problem {\it iii}.
We choose a middle ground and try to identify
as many properties as we can that have some plausible association with habitability.
This strategy is most sensitive if a 
few unknown stellar properties (among the ones being tested) contribute to the 
habitability of a terrestrial planet in orbit around a star.

An optimal quantitative comparison of the Sun to other stars would require an unbiased, large 
representative stellar sample from which independent distributions, for as many 
properties as desired,  could be compared. Such a distribution for each 
property of interest would allow a straightforward 
analysis and outcome:  the Sun is within the $n\%$ of stars around the centroid of the 
$N$-dimensional distribution. However, observational and sample selection effects prevent 
the assembly of such an ideal stellar sample. 

In this study, we compare the Sun to other stars with respect to the following 11 basic
physical properties: 
(1) mass, (2) age, (3) metallicity [Fe/H], 
(4) carbon-to-oxygen ratio [C/O],  (5) magnesium-to-silicon ratio [Mg/Si], 
(6) rotational velocity $v\sin i$, (7) eccentricity of the star's galactic orbit $e$, 
(8) maximum height to which the star rises above the galactic plane $Z_{\mr{max}}$,
(9) mean galactocentric radius R$_{\mr{Gal}}$, 
(10) the mass of the star's host galaxy M$_{\mr{gal}}$,
(11) the mass of the star's host group of galaxies M$_{\mr{group}}$.
These 11 properties span a wide range of stellar and galactic factors that may be 
associated with habitability. We briefly discuss how each parameter might have a plausible 
correlation with habitability. For each property we have tried to assemble a large, 
representative sample of stars whose selection criteria is minimally biased with respect to that property.
For each property the percentage of stars with values lower and higher than the solar value are computed. 
For properties (9), (10) and (11), the uncertainties in the percentages are determined from the 
uncertainties of the distributions. For the rest of the properties, nominal 
uncertainties $\Delta$, on the percentages were calculated assuming a binomial distribution 
(e.g.\ \citealt{meyer75}): $\Delta =(n_{\mr{low}} \times n_{\mr{high}} / N_{\mr{tot}})^{1/2}$ 
where $n_{\mr{low}}$ ($n_{\mr{high}}$) is the fraction of stars with a lower (higher) value 
than the Sun and $N_{\mr{tot}}$  is the total number of stars in the sample. The solar value 
is indicated with the symbol ``$\odot$'' in all figures.

We compare the Sun and its environment to other stars and their environments.  The analysis 
of these larger environmental contexts provides information about properties that otherwise 
could not be directly measured. For example, suppose the metallicity of the Sun were 
normal with respect to stars in the solar neighborhood but that these stars as a group, had an 
anomalously high metallicity with respect to the average metallicity of stars in the Universe. 
This fact would strongly suggest that habitability is associated with high metallicity, but our 
comparison with only local stars would not pick this up. In the absence of an [Fe/H] distribution
for all stars in the Universe, we use galactic mass as a convenient proxy for any such property that 
correlates with galaxy mass.

%%%%%%%%%%  fig: mass-histogram
\begin{figure}[!hbt]
  \begin{center}
    \includegraphics[scale=0.55]{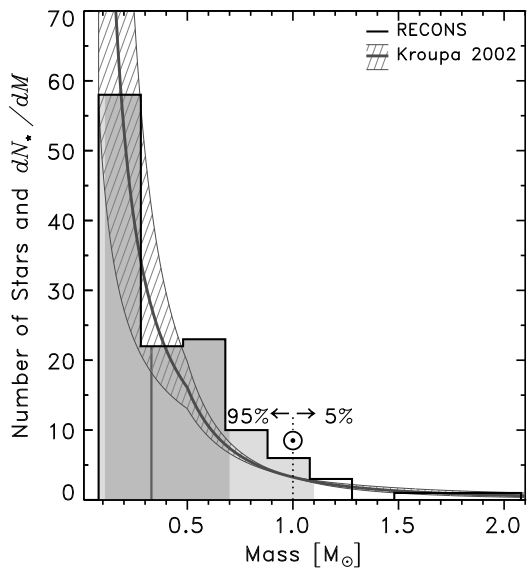}
    \caption{Mass histogram of the 125 nearest stars \citep[][RECONS]{Henry06}. 
      The median ($\mu_{1/2} = 0.33$ M$_{\odot}$) of the distribution is indicated by the vertical 
      grey line. The 68\% and 95\% bands around the median are indicated respectively by the vertical 
      dark grey and light grey bands. We also use these conventions in Figs.\ 2--11. The solid 
      curve and hashed area around it represents the Initial Mass Function (IMF) and its associated uncertainty 
      \citep{Kroupa02}. The Sun, indicated by ``$\odot$'', is more massive than $95 \pm2\%$  of these stars.}
    \label{fig:mass}
  \end{center}
\end{figure}
%%%%%%%%%%  end fig: mass histogram 

\subsection{Mass} \label{sec:mass}
Mass is probably the single most important characteristic of a star. For a main sequence 
star, mass determines luminosity, effective temperature,  main sequence life-time and 
the dimensions, UV insolation and temporal stability of the circumstellar habitable 
zone \citep{kasting93}. 

Low mass stars are intrinsically dim.  Thus a complete sample
of stars can only be obtained out to a distance of $\sim 7$ parsecs ($\approx 23$ lightyears).
Figure \ref{fig:mass} compares the mass of the Sun to the stellar mass 
distribution of the 125 nearest main sequence stars within 7.1 pc, as compiled by the 
RECONS consortium \citep{Henry06}.
Over-plotted is the stellar Initial Mass Function \citep[IMF; see eqs.~4 \& 5 and Table~1 of][]{Kroupa02} 
normalised to 125 stars more massive than the brown dwarf limit of $0.08 \; \mr{M}_{\odot}.$
Since the IMF appears to be fairly universal \citep{Kroupa05}, these nearby comparison stars 
are representative of a much larger sample of stars.
There is good agreement between the histogram and the IMF --- the Sun is more massive 
than $95 \pm2\%$ of the nearest stars, and more massive than $94 \pm2\%$ of the stars in the 
\citet{Kroupa02} IMF. Fourteen brown dwarfs and nine white dwarfs within 7.1 pc were not 
included in this sample. Including them yields $94\%$ --- the same result obtained from the IMF.
Our $95\% \pm 2\%$ result should be compared with the $91\%$ reported by \cite{Gonzalez99b}.
The Sun's mass is the most anomalous of the properties studied here.

\subsection{Age} \label{sec:age}
If the evolution of observers like ourselves takes on average  many billions of years, 
we might expect the Sun to be anomalously old \citep{Carter83}.
Accurate estimation of stellar ages is difficult. For large stellar surveys ($>$ a few hundred stars), 
the most commonly used age indicators are based on isochrone fitting and/or chromospheric  
activity ($R'_{\mr{HK}}$ index).
\cite{Rochapinto00a} have estimated a Star Formation Rate (SFR), or equivalently, an age distribution
for the local Galactic disk from chromospheric ages of 552 late-type (F8--K2) dwarf stars in the 
mass range $0.8 \; \mr{M}_{\odot} \le M \le 1.4 \; \mr{M}_{\odot}$ at distances $d \le 200$ pc 
\citep{Rochapinto00b}. They applied scale-height corrections, stellar evolution corrections 
and volume incompleteness corrections that converted the observed age distribution into the 
total number of stars born at any given time. \cite{hernandez00} and \cite{bertelli01} have
made estimates of the star formation rate in the solar neighborhood and favour a smoother
distribution (fewer bursts) than \cite{Rochapinto00a}.

\begin{figure}[!hbt]
        \begin{center}
                \includegraphics[scale=0.55]{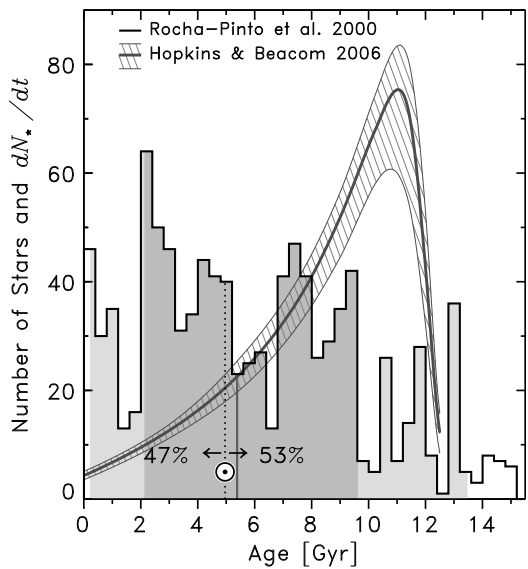}
                \caption{The Galactic stellar age distribution (median $\mu_{1/2} = 5.4$ Gyr) from 
\cite{Rochapinto00a}. The Sun is younger than $53 \pm 2 \%$ of the stars in the disk 
of our Galaxy. The grey curve is the cosmic Star 
Formation Rate (SFR) with its associated uncertainty \citep{Hopkins06}, according to 
which the Sun is younger than $86 \pm 5 \%$ of the stars in the Universe.}
                \label{fig:age}
        \end{center}
\end{figure}

In Figure \ref{fig:age} we compare the chromospheric age of the Sun 
\citep[$\tau_{\odot}=4.9 \pm 3.0$ Gyr,][]{Wright04}
\footnote{To ensure that the Sun's age is determined in the same way as the stellar ages to which 
it is being compared, we adopt the chromospheric solar age $\tau_{\odot}= 4.9 \pm 3.0$ Gyr 
over the more accurate meteoritic age $\tau_{\odot}=4.57\pm0.002$ Gyr \citep{Allegre95}.} 
to the stellar age distribution representing the Galactic SFR \citep{Rochapinto00a}. 
The median of this distribution is 5.4 Gyr. The Sun is younger than $53 \pm 2 \%$ of the stars 
in the thin disk of our Galaxy. Over-plotted is the cosmic SFR derived by \cite{Hopkins06}. 
According to this distribution with a median $\mu_{1/2} = 9.15$ Gyr, the Sun was born after 
$86 \pm 5 \%$ of the stars that have ever been born.

The Galactic and cosmic SFRs are different because the cosmic SFR 
was dominated by bulges and elliptical galaxies in which the largest
fraction of stellar mass in the Universe resides.
Bulges and elliptical galaxies (early-type galaxies) formed their stars early and quickly and then ran out
of gas. The disks of spiral galaxies, like our Milky Way, seem to have undergone irregular bursts of star formation
over a longer period of time as they interacted with their satellite galaxies. 

The volume limited ($d_{\mr{max}}=$ 40 pc) sub-set from \cite{Nordstrom04} contains isochrone ages for 
1126 A5--K2 stars. The median of this sub-set is 5.9 Gyr and the Sun is younger than $55 \pm 2\%$ 
of the stars. The similarity of this isochrone age result to the chromospheric age result 
is not obvious since the agreement between these two age techniques is rather poor.
This mismatch can be seen in Fig.~\ref{fig:massagev}D, \cite{reid07}, and
in Fig.~8 of \citet{Feltzing01}.

%%%%%%%%%%  fig: age-histogram
%%%%%%%%%%  end fig: age histogram 

%METALLICITY
\subsection{Metallicity} \label{sec:metallicity}
Iron is the most frequently measured element in nearby stars.  
Metallicity [Fe/H], is known to be a proxy for the 
fraction of a star's mass that is not hydrogen or helium.
In the Sun and possibly in the Universe, the dominant contributors to this mass fraction in 
order of abundance are: O(44\%), C(18\%), Fe(10\%), Ne(8\%), Si(6\%), Mg(5\%), N(5\%), 
S(3\%) \citep{Asplund05,Truran05}. The corresponding abundances by number are: O(48\%), 
C(26\%), Ne(7\%), N(6\%), Mg (4\%), Si(4\%), Fe(3\%), S(2\%). Importantly for this analysis, 
this short list contains the dominant elements in the composition of terrestrial 
planets (O, Fe, Si and Mg) and life (C, O, N and S).

Over the last few decades, much effort has gone into determining abundances in nearby stars 
for a wide range of elements. Stellar elemental abundances for element X are usually normalised to 
the solar abundance of the same element using a logarithmic abundance scale:  
\mbox{$\mr{[X/H]_{\star} \equiv \log(X/H)_{\star} - \log(X/H)_{\odot}}$}. Hence all solar elemental 
abundances [X/H]$_{\odot}$, are defined as zero. Spectroscopic abundance analyses are usually made 
differential relative to the Sun by analysing the solar spectrum (reflected by the Moon, asteroids 
or the telescope dome) in the same way as the spectrum of other stars. In this approach, biases 
introduced by the assumption of local thermodynamic equilibrium (LTE), largely cancel out for Sun-like
stars \citep{Edvardsson93b}.

A comparison between solar and stellar iron abundances is a common feature of most abundance surveys
and most have concluded that the Sun is metal-rich compared to other stars 
\citep{Gustafsson98, Gonzalez99a, Gonzalez99b}. However, for our purposes,  the appropriateness of 
these comparisons depends on the selection criteria of the stellar sample to which the Sun has been compared. 
Stellar metallicity analyses such as \cite{Edvardsson93, Reddy03, Nordstrom04, Valenti05}
have stellar samples selected with different purposes in mind, e.g., \cite{Edvardsson93} aimed to constrain the 
chemical evolution of the Galaxy and their sample is biased towards low metallicity (average [Fe/H]$=-0.25$).   
The sample of \citealt{Valenti05} (average \mbox{[Fe/H]$=-0.01$}), was selected as a planet candidate list and 
contains some bias towards high metallicity \citep[see][]{Grether07}. To assess how typical the Sun is, 
\cite{Gustafsson98} limited the sample of  \cite{Edvardsson93} to stars with galactocentric radii within 
0.5 kpc of the solar galactocentric radius, and to ages between 4 and 6 Gyrs. The distribution of stars 
given by this criteria has an average [Fe/H]$=-0.09$. 

\begin{figure}[!t]
        \begin{center}
                \includegraphics[scale=0.55]{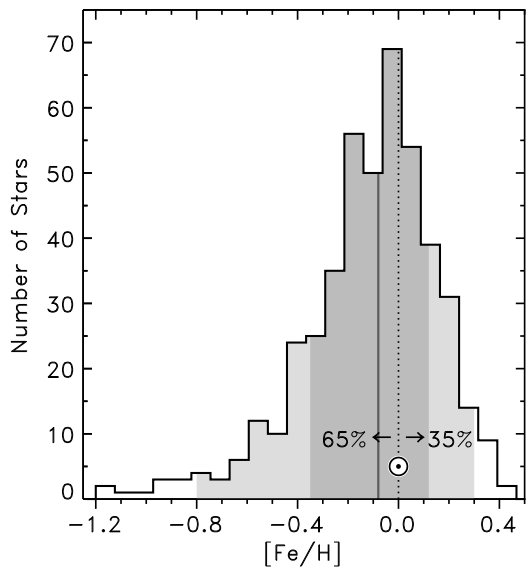}
                \caption{Stellar metallicity histogram of the 453 FGK \textit{Hipparcos} stars 
                within 25 pc \citep{Grether07}.  The median $\mu_{1/2} = -0.08$. 
                The Sun is more metal-rich than $65 \pm 2\%$  of the stars.}
                \label{fig:fe}
        \end{center}
\end{figure}

\cite{Grether06,Grether07} compiled a sample of 453 Sun-like stars within 25 pc. 
These stars were selected from the \textit{Hipparcos} catalogue, which is essentially complete to
25 pc  for stars within the spectral type range F7--K3  and 
absolute magnitude of $M_V=8.5$ \citep{Reid02}.
Metallicities for this sample were assembled from a wide range of spectroscopic and photometric surveys.
In Figure \ref{fig:fe}, we compare the Sun to the \cite{Grether07} sample, which has a median [Fe/H]$=-0.08$.
To our knowledge this is the most complete and least-biased stellar spectroscopic metallicity distribution. 
The Sun is more metal-rich than $65 \pm2\%$ of these stars.

This result should be compared with \cite{Favata97} who constructed a volume-limited  
($d_{\mr{max}} =$ 25 pc) sample of 91 G and K dwarfs ranging in color index ($B-V$) 
between $0.5-0.8$ (\citealt{Favata96}). Their distribution has a median [Fe/H]$=-0.05$ 
and compared to this sample, the Sun is more metal rich than $56 \pm 5 \%$  of the stars.
\cite{fuhrmann08} compared the Sun to a volume complete ($d_{\mr{max}} =$ 25 pc) sample of about
185 thin-disk mid-F-type to early K-type stars down to $M_V = 6.0$. He finds a mean
[Fe/H] = $-0.02 \pm 0.18$. This mean [Fe/H] is lowered by 0.01 dex if the 43 double-lined
spectroscopic binaries in his sample are included. His results are consistent with ours.

%%%%%%%%%%  fig: fe-histogram
%%%%%%%%%%  end fig: fe histogram 

\subsection{Elemental Ratios $\mathrm{[C/O]}$ and $\mathrm{[Mg/Si]}$}\label{sec:ratios}
The elemental abundance ratios of a host star have a major impact on its proto-planetary disk chemistry and 
the chemical compositions of its planets. Oxygen and carbon make up $\sim 62\% $ of the solar system's 
non-hydrogen-non-helium mass content \citep[$Z=0.0122$,][]{Asplund05}. 
Carbon and oxygen abundances are among the hardest to determine.
This is due to high temperature sensitivity and non-LTE effects in their permitted lines 
(e.g.\ \ion{C}{1} $\lambda6588$, \ion{O}{1} $\lambda7773$), and to the presence of blends 
in the forbidden lines ([\ion{C}{1}] $\lambda8727$, [\ion{O}{1}] $\lambda6300$). 
See \cite{allendeprieto01} and \cite{Bensby06} for details on C and O abundance derivations.

Carbon pairs up with oxygen to form carbon monoxide. In stars with a C/O ratio larger than 1,
most of the oxygen condenses into CO which is largely driven out of the incipient circumstellar 
habitable zone by the stellar wind. In this oxygen-depleted scenario, planets formed within the snow line are 
formed in reducing environments and are mostly composed of carbon compounds, for example silicon 
carbide \citep{Kuchner05}. Thus, the C/O ratio could be strongly associated with habitability.

As most heavy element abundances relative to hydrogen (e.g.\ [O/H], [C/H],
[N/H]) are correlated with [Fe/H], they were not included in our analysis.
After the overall level of metallicity (represented by [Fe/H]), and after the ratio of the two most abundant 
metals, [C/O], the magnesium to silicon ratio [Mg/Si] is the most important ratio of the next most abundant 
elements (excluding the noble gas Ne).  For example [Mg/Si] sets the ratio of olivine to pyroxene which 
determines the ability of a silicate mantle to retain water (H. O'Neill 2007, private communication).

Stellar elemental abundance ratios are defined as $\mr{ [X_1/X_2]_{\star}=[X_1/H]_{\star} - [X_2/H]_{\star} }$. 
Hence, systematic errors associated with the determination of absolute solar abundances cancel for 
abundances relative-to-solar. We compile [C/O] and [Mg/Si] ratios from samples with the 
largest number of stars and highest signal-to-noise ratio stellar spectra:

\begin{description}
\item{[C/O].}---256 stars from  \citealt{Gustafsson99,Reddy03,Bensby06} %(G99, R03, BF06)
\item{[Mg/Si].}---231 stars from \citealt{Reddy03,Bensby05} %(R03,B05)
\end{description}

%%%%%%%%%%  fig: co,mg-histogram
\begin{figure*}[!t]
     \begin{center}
     \begin{tabular}{rr}
        \includegraphics[scale=0.55]{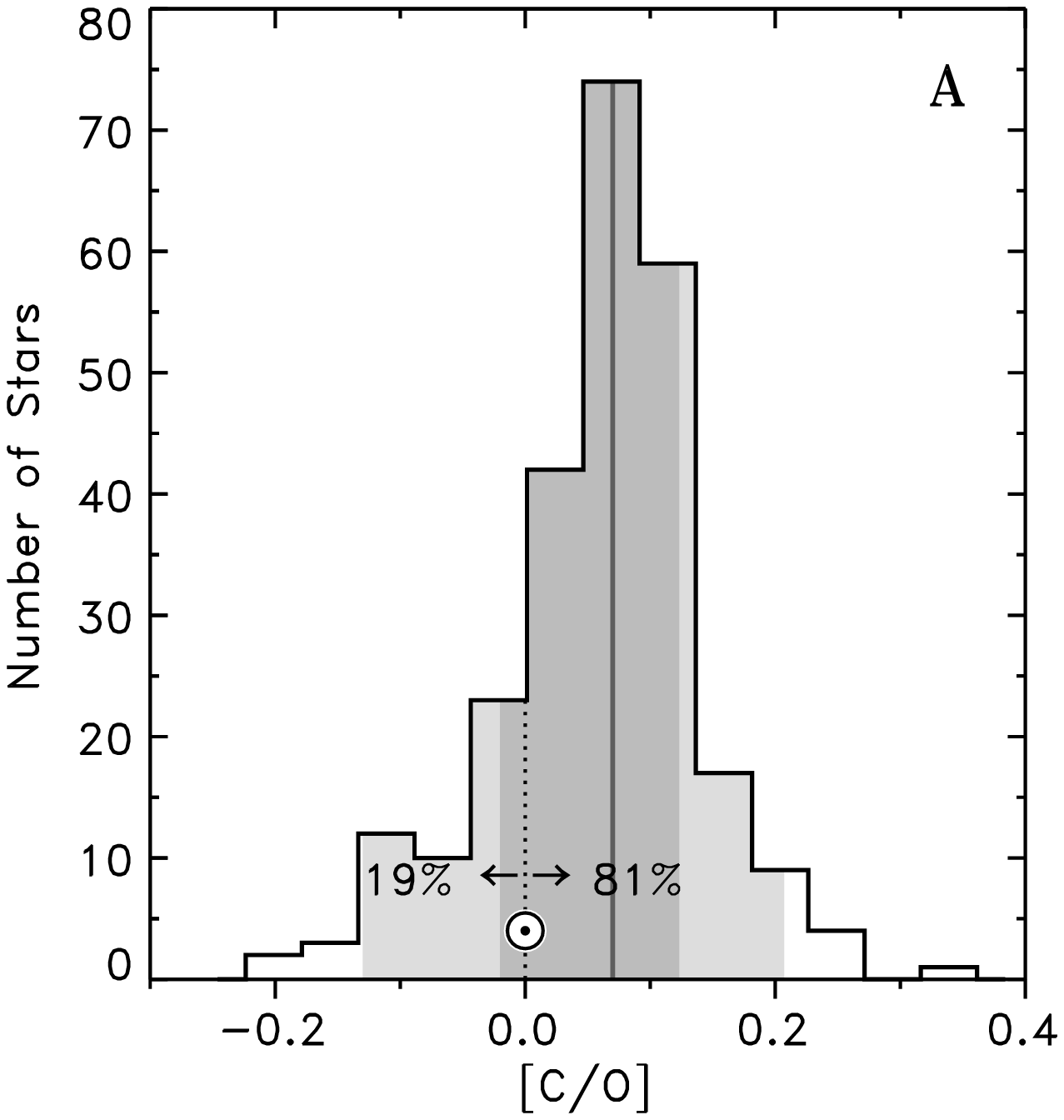} & \includegraphics[scale=0.55]{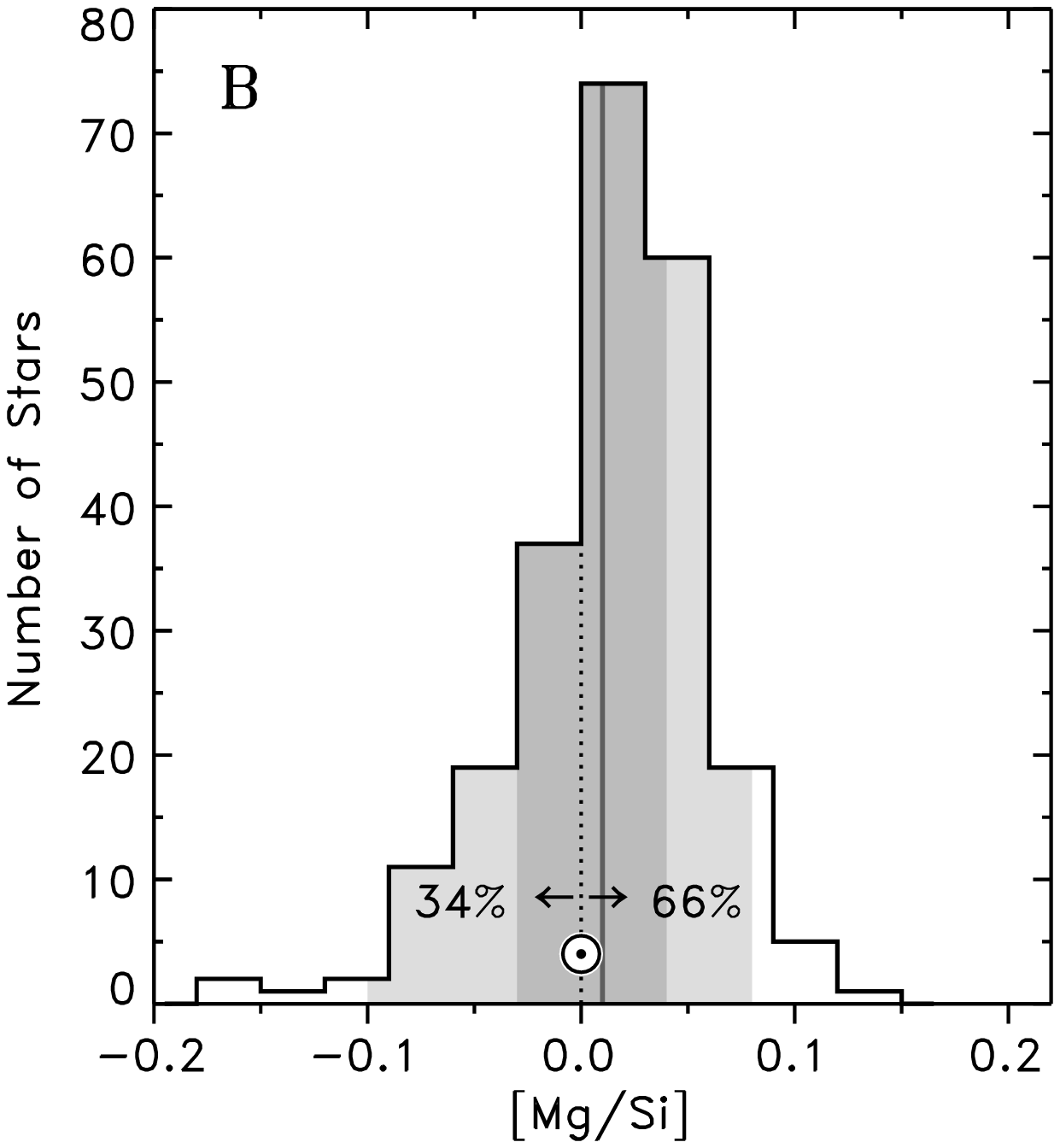} \\
        \includegraphics[scale=0.55]{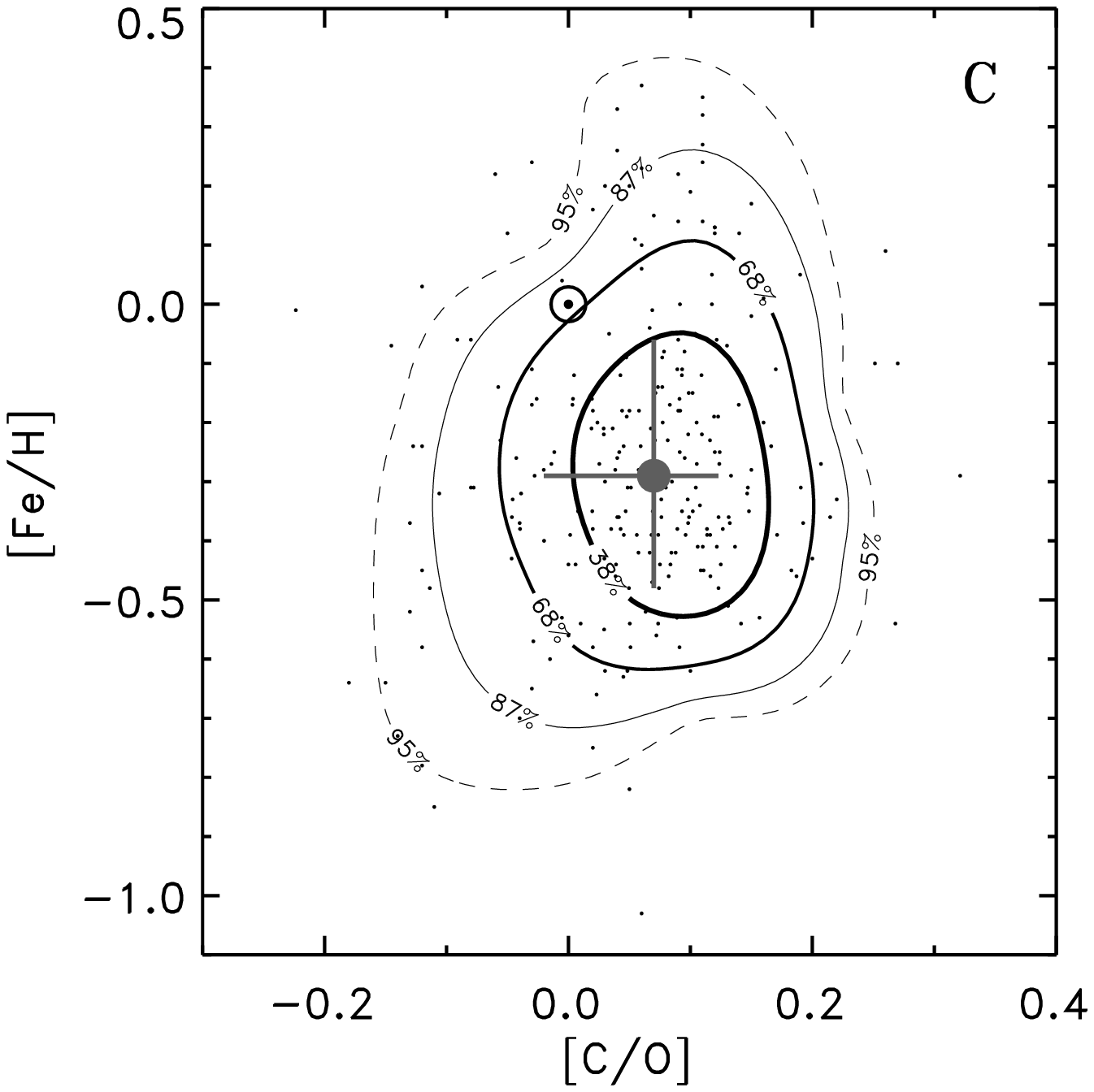} & \includegraphics[scale=0.55]{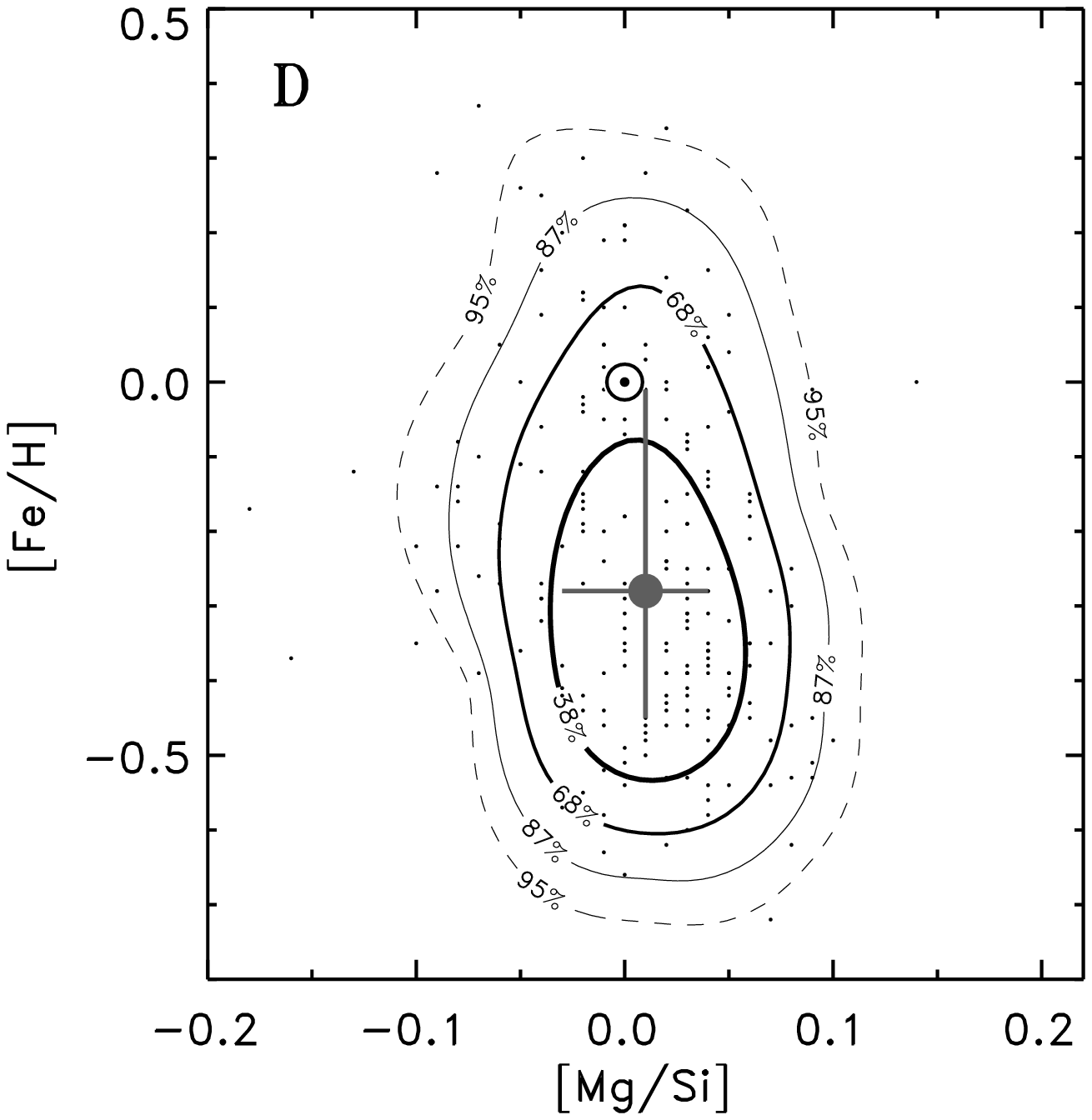} \\
     \end{tabular}
                \caption{ {\bf A}: Comparison of the Sun's carbon-to-oxygen ratio ([C/O]$_{\odot} \equiv 0$) 
                  to the [C/O] ratios of 256 stars compiled from \cite{Gustafsson99,Reddy03} and \cite{Bensby06}.
                  The Sun's [C/O] ratio is lower than $81 \pm3 \%$ of the stars in this sample which has a 
                  median $\mu_{1/2}=0.07$. 
                  {\bf B}: Comparison of the Sun's magnesium-to-silicon ratio  ([Mg/Si]$_{\odot} \equiv 0$), 
                  to [Mg/Si] values from 231 stars from \citet{Reddy03} and \citet{Bensby05}. 
                  The Sun's [Mg/Si] ratio is lower than $66 \pm3 \%$ of the stars in this sample 
                  with median $\mu_{1/2}=0.01$. The bottom panels {\bf C} \& {\bf D} show the small 
                  correlations of these distributions with [Fe/H]. These small correlations can be 
                  neglected for this study.
                }
                \label{fig:ratios}
        \end{center}
\end{figure*}
%%%%%%%%%%  end fig: co,mg-histogram
Due to their selection criteria, these samples are biased towards low metallicity and therefore 
cannot be used to create a representative [Fe/H] distribution. Because a correlation exists between
the [C/O] and [Mg/Si] ratios and [Fe/H] \citep[e.g.][]{Gustafsson99}, the
samples we use have a relatively narrow range of [Fe/H] to reduce the influence of the correlation.
Therefore, these small correlations can be neglected in this study --- see the bottom panels of Fig.~\ref{fig:ratios} 
where [Fe/H] versus [C/O] as well as [Fe/H]  versus [Mg/Si] are plotted. 
The top panels show the corresponding stellar distribution histograms. The Sun's [C/O] 
ratio is lower than $81 \pm3 \%$ of the stars.
This is consistent with \cite{Gonzalez99b} who suggested --- based on data from \cite{Edvardsson93} 
and \cite{Gustafsson99}  --- that the Sun has a low [C/O] ratio
relative to Sun-like stars  at similar galactocentric radii. 
See however, \cite{ramirez07} who find that the Sun is oxygen poor
compared to solar metallicity stars. 

The Sun's [Mg/Si] ratio is lower than $66 \pm3 \%$ of the stars. The [C/O] and [Mg/Si] ratios 
are also largely independent of each other (see Fig.~\ref{fig:comgsi} in Appendix A).

\subsection{Rotational Velocity} \label{sec:rotvel}
Stellar rotational velocities are related to 
the specific angular momentum of a protoplanetary disk and possibly to 
the magnetic field strength of the star during planet formation, and  to protoplanetary disk 
turbulence and  mixing. An unusually low stellar rotational velocity may be associated with 
the presence of planets \citep{soderblom83}. One or several of these factors could be 
related to habitability.

There is a known correlation between mass and $v\sin i$
at higher stellar masses  \citep[e.g.\ see Fig.~18.21 of][p.~485]{gray05}. In order to minimise 
the effect of this correlation (and maximize independence between parameters), we assembled 
a sample containing 276 stars within the mass range 0.9--1.1 M$_{\odot}$ (F8--K2) from 
\cite{Valenti05}. The selection criteria of the \cite{Valenti05} stars 
introduces some bias against more active stars. We compared 
the high $v\sin i$ tail of our \cite{Valenti05} sample with the high $v\sin i$ tail of a sub-sample 
from \cite{Nordstrom04}.  We estimate that for our \cite{Valenti05} sample, the bias introduced by 
the selection criteria is lower than $\sim 5\%$. The $v\sin i$ values in \cite{Valenti05}
are obtained by fixing the macroturbulence for the stars of a given color, without modeling 
the stars individually. If the macroturbulence value was underestimated for $T>5800$ K, 
the resulting $v\sin i$ values (especially when $v\sin i$ is near zero) would be 
overestimated \citep[Sec.~4 of][]{Valenti05}.

The inclination of the stellar rotational axis to the line of sight is usually unknown so the
observable is $v\sin i$. Using the solar spectrum reflected by the asteroid Vesta, \cite{Valenti05} derived a 
solar $v\sin i = 1.63$ \mbox{km s$^{-1}$}. For the purposes of this analysis we use the mean value that 
would be derived for the Sun, when viewed from a random inclination:
$v\sin i_{\odot} = 1.63 (\pi /4)$ \mbox{km s$^{-1}$} $\approx 1.28$ \mbox{km s$^{-1}$}.

\begin{figure}[!t]
  \begin{center}
    \includegraphics[scale=0.55]{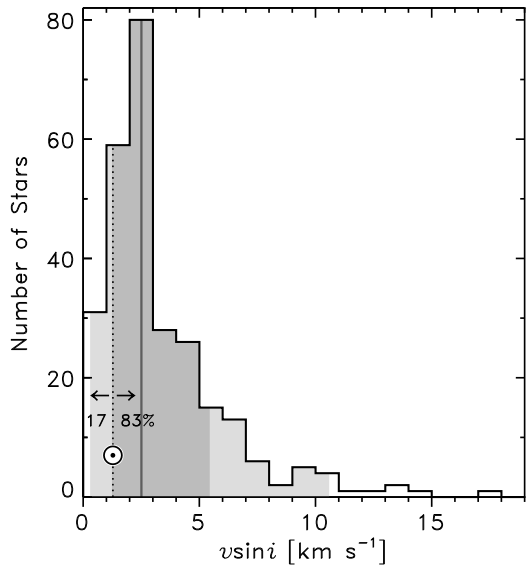}
    \caption{Rotational velocity histogram for 276 F8--K2 ($0.9 \leq M \leq 1.1$ M$_{\odot}$) 
      stars \citep{Valenti05}. 
      The Sun ($v\sin i_{\odot} = 1.28$ \mbox{km s$^{-1}$}) rotates more slowly than $83 \pm 7\%$ of the stars.
      There is 1 star to the right of the plot with $v\sin i = 36$ \mbox{km s$^{-1}$}.
    }
    \label{fig:vsini}
  \end{center}
\end{figure}

The Sun rotates more slowly than $83 \pm 7\%$ of the stars in our \cite{Valenti05} 
sample (Fig.~\ref{fig:vsini}). This is in agreement with \cite{soderblom83,soderblom85} 
who reported that the Sun is within 1 standard deviation of stars of its mass and age.

%%%%%%%%%%  Fig: vsini histogram
%%%%%%%%%%  end fig: vsini histogram 

%%%%%%%%%%  fig: e
\begin{figure}[!h]
  \begin{center}
    \begin{tabular}{c}
      \includegraphics[scale=0.50]{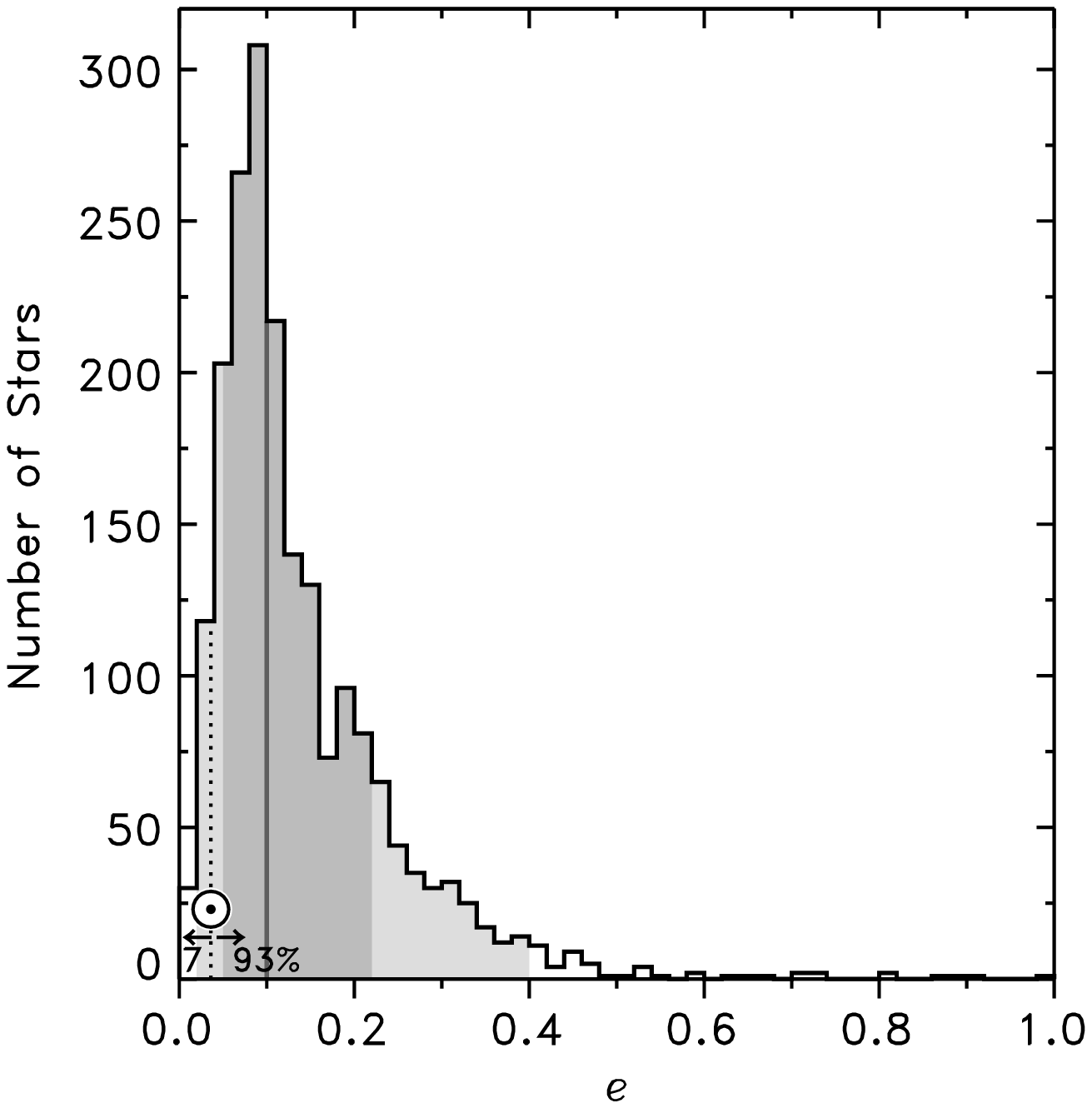} \\
      \includegraphics[scale=0.50]{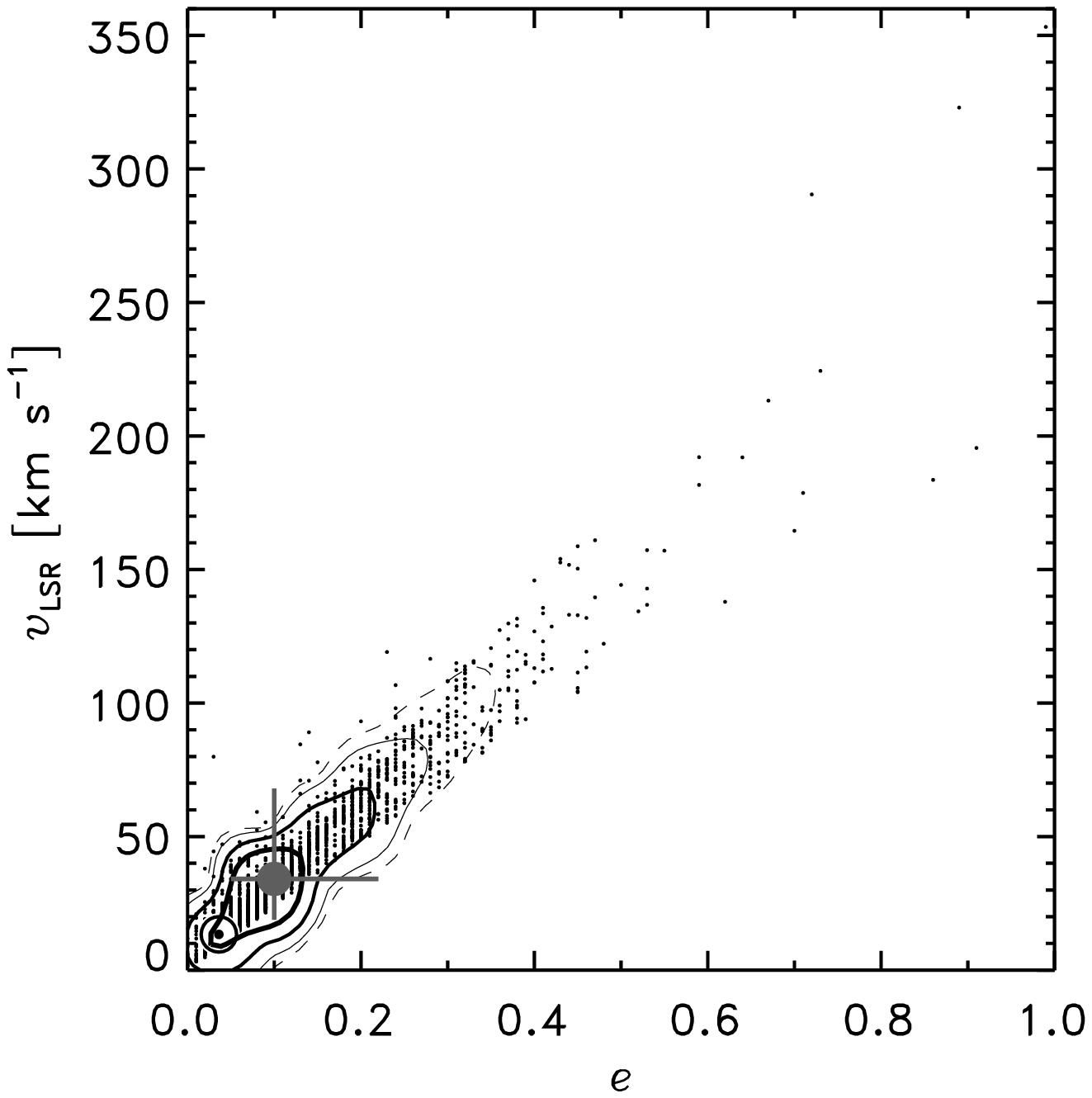}
    \end{tabular}
    \caption{Top: eccentricity distribution for the 1,987 stars at $d\leq 40$ pc from 
      \cite{Nordstrom04}. The Sun has a more circular orbit than $93 \pm1 \%$ 
      of the A5--K2 stars within 40 pc. After mass, eccentricity is the second most anomalous parameter.
      Bottom panel: Correlation between $v _{\mr{LSR}}$ and eccentricity for the same stars presented in the 
      Bottom: since these properties are highly correlated we select only 1 for the analysis. 
      The large grey point with error bars represents the median and the $68\%$ widths of the
      two one-dimensional distributions. As in Fig.~\ref{fig:ratios} the contours 
      correspond to 38\%, 68\%, 82\% and 95\%.
           }
    \label{fig:e}
   \end{center}
\end{figure}
%%%%%%%%%%  end fig: e

\subsection{Galactic Orbital Parameters} \label{sec:orbital}
The Galactic velocity components of a star ($U$,$V$,$W$) with respect to the local standard of rest (LSR)
may be used to compute a star's orbit in the Galaxy.
How typical or atypical is the solar orbit compared to the orbits of other nearby stars in the Galaxy? 
The orbit may be related to habitability because more eccentric orbits
bring a star closer to the Galactic center where there is a larger danger to life from
supernovae explosions, cosmic gamma and X-ray radiation and any factors associated with higher 
stellar densities \citep{Gonzalez01,Lineweaver04}. 

For a standard model of the Galactic potential, 
\cite{Nordstrom04} computed 
orbital paramters for
the Sun, and for a large sample ($\sim 16700$) of A5--K2 stars.  
Their adopted components of the solar velocity relative to the local
standard of rest were
$(U,V,W)=(10.0 \pm 0.4, 5.25\pm 0.62, 7.17\pm 0.38)$ \mbox{km s$^{-1}$} \citep{dehnen98}.

For each of the 1987 stars within 40 pc in the \cite{Nordstrom04} catalog, 
an inner and outer radii $R_{\mr{min}}$ and
$R_{\mr{max}}$ were computed.  This yielded the orbital 
eccentricity $e \equiv (R_{\mr{max}}-R_{\mr{min}})/(R_{\mr{min}}+R_{\mr{max}})$.  
The solar eccentricity was computed using the components of the 
solar motion \citep{dehnen98} relative to the local standard of rest in the 
Galactic potential of \cite{flynn96}.
The bottom panel of Figure \ref{fig:e} shows the correlation between Galactic
orbital eccentricity $e$ and the magnitude of the galactic orbital velocities
with respect to the local standard of rest:
$v_{\mr{LSR}}\equiv (U^2+V^2+W^2)^{1/2}$. 
Eccentricity $e$ and $v_{\mr{LSR}}$ are strongly correlated. We include $e$, not $v_{\mr{LSR}}$, 
in the analysis since $e$ is less correlated with the maximum height above the
Galactic plane $Z_{\mr{max}}$, than is $v_{\mr{LSR}}$. This is shown in Fig.~\ref{fig:ezmax} in 
Appendix A.

The Sun's eccentricity was determined with the same relation as the stellar 
eccentricities. The uncertainty in our estimate of solar eccentricity came from
propagating the uncertainty in the adopted solar motion. We find
$e_{\odot}=0.036\pm 0.002$ \citep[consistent with the 
$e_{\odot}=0.043\pm 0.016$ found by][]{Metzger98}.
The Sun has a more circular orbit than $93 \pm1 \%$ of the A5--K2 stars within 
40 pc (with median eccentricity $\mu_{1/2}=0.1$).
This is the second most anomalous of the 11 solar properties we consider here.

\begin{figure}[!ht]
   \begin{center}
      \includegraphics[scale=0.55]{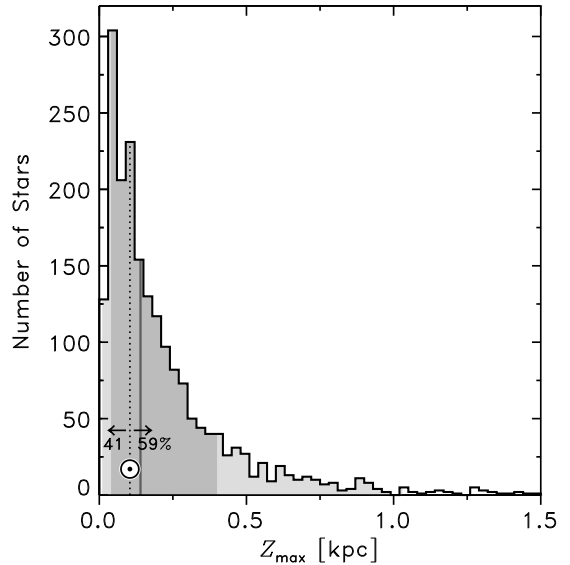}
      \caption{The distribution of maximum heights above the Galactic plane for the \cite{Nordstrom04} sample. 
               $59 \pm 3\%$ of nearby A5--K2 
               stars ($d_{\mr{max}}=$40 pc) reach higher above the Galactic plane than the Sun reaches.
               There are 22 stars evenly distributed  
               over $Z_{\mathrm{max}}$ between 1.5 and 9.6 kpc. Their exclusion
               from the comparison reduces the 59\% result by less than 1\%.
             }
      \label{fig:z}
   \end{center}
\end{figure}

The frequency of the passage of a star through the thin disk could be associated with 
Galactic gravitational tidal perturbations of Oort cloud objects that might increase 
the impact rate on potentially habitable planets. This is correlated with the maximum 
height, $Z_{\mr{max}}$, to which the stars rise above the Galactic plane.
Figure \ref{fig:z} shows the stellar distribution of $Z_{\mr{max}}$ for the stars shown 
in Figure \ref{fig:e}. We find that $59 \pm 3\%$ of the A5--K2 stars
within 40 pc of the Sun reach higher above the Galactic plane than the 
Sun does ($Z_{\mr{max},\odot}=0.104 \pm 0.006$ kpc). The solar  $Z_{\mr{max},\odot}$ was 
derived by integrating the solar orbit in the Galactic potential. The
uncertainty on $W$, produces the uncertainty on $Z_{\mr{max},\odot}$
and hence the $\pm 3\%$ uncertainty on $59\%$. Our results for eccentricity
and $Z_{\mr{max}}$ are consistent with those obtained using \cite{hogg05} LSR values:
$(U,V,W) = (10.1 \pm 0.5, 4.0 \pm 0.8, 6.7 \pm 0.2)$. Using the Hogg et al. LSR values, 
$92 \pm 1\%$ of A5--K2 stars within 40 pc have higher eccentricities than the Sun and 
$62 \pm 4\%$ of A5--K2 stars within 40 pc have larger $Z_{\mr{max}}$ values.

\begin{figure}[!htb]
   \begin{center}
       \includegraphics[scale=0.55]{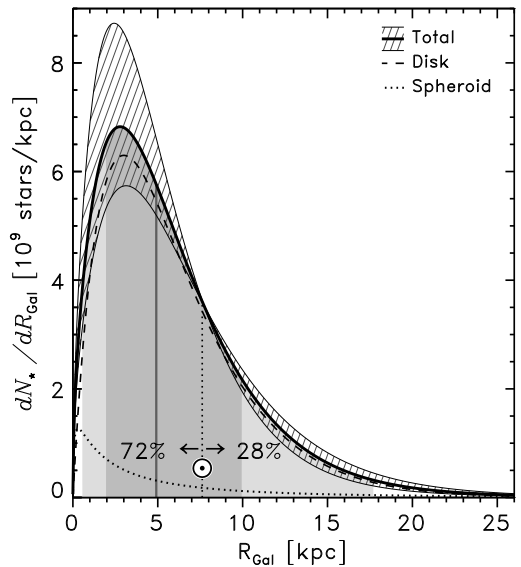} 
       \caption{Mean stellar galactocentric radius distribution $dN_{\star}/dR_{\mr{Gal}}$. The solid curve 
         represents the sum of the disk (dashed line) and spheroidal (dotted line) stellar components.
         The $68\%$ uncertainty of the total distribution is shown by the cross-hatched area. 
         The Sun is farther from the Galactic center than $72^{+8}_{-5}\%$ of the stars in the Galaxy.}
       \label{fig:radius}
   \end{center}
\end{figure}

How does the Sun's distance from the center of the Milky Way
compare to the distances of other stars from the center of the Milky Way?
In Fig.~\ref{fig:radius} we show the distribution of the mean radial distances of stars from the Galactic 
center, based on the star count model of \cite{Bahcall80}. To represent the entire Galactic stellar 
population we include the disk (thin + thick) and spheroidal (bulge + halo) components.
Using the current Solar distance from the center \citep[$R_0 = 7.62 \pm 0.32$ kpc,][]{Eisenhauer05} 
and a disk scale length $h=3.0\pm0.4$ kpc \citep{Gould96}, we estimate that the Sun lies farther from 
the Galactic center than $72^{+8}_{-5}\%$ of the stars in the Galaxy. The uncertainty on the 
result comes from the 68\% bounds of the total 
distribution, which come from the scale length uncertainty ($\pm 0.4$ kpc).

%%%%%%%%%%  fig: z
%%%%%%%%%%  end fig: z 

%%%%%%%%%%  fig: radius-distribution
%%%%%%%%%%  end fig: radius distribution

\newpage

 \subsection{Host Galaxy Mass}     \label{sec:galaxymass}
The mass of a star's host galaxy may be correlated with parameters that have an 
influence on habitability. For example, galaxy mass affects the overall metallicity 
distribution that a star would find around  itself --- an effect that would not show up in 
Fig.~\ref{fig:fe}, which only shows the local metallicity distribution.

\begin{figure}[!t]
  \begin{center}  
    \includegraphics[scale=0.55]{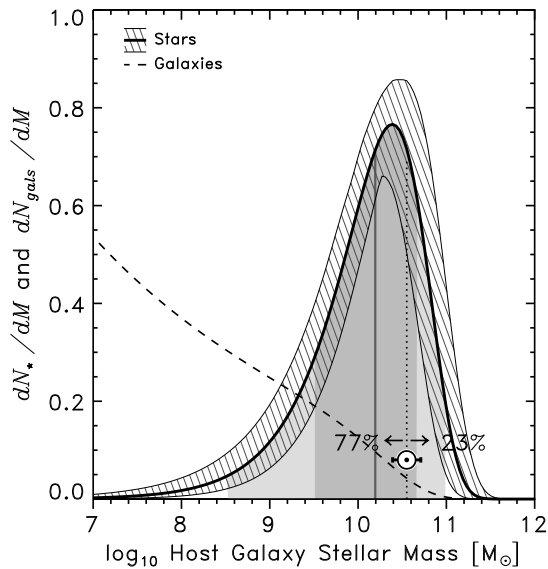}
    \caption{
      Fraction of all stars that live in galaxies of a given mass, $dN_{\star}/dM$ (solid curve).
      The mass of the Sun's galaxy is indicated by the ``$\odot$''.
      This distribution represents the amount of stellar mass contributed by galaxies of a given mass. 
      Approximately $77^{+11}_{-14} \%$ of stars live in galaxies less massive than ours.
      The cross-hatched band shows the $1\sigma$ uncertainty associated with the uncertainty in the two 
      Schechter function parameters,  $\alpha$ and $L^{*}$ \citep{lovedayj00,schechter76}. 
      The dashed line shows the unweighted luminosity function (the number of galaxies per luminosity 
      interval $dN_{gals}/dM$) according to which the Milky Way is more massive than $\sim 99\%$ of galaxies.
    }
    \label{fig:galaxy}
  \end{center}
\end{figure}

The Milky Way is more massive than $\sim 99$\% of all galaxies --- the precise fraction 
depends on the lower mass-limit chosen for an object to be classified as a galaxy, and the 
behaviour of the low-mass end of the galaxy mass function \citep{silk07}.  We are referring 
here to the stellar mass, not the total baryonic mass or the total mass.  
Despite the Milky Way's large mass compared to other galaxies, if most stars in the Universe 
resided in even more massive galaxies, the Milky Way would be a rather low mass galaxy for a 
star to belong to. To estimate the fraction of all stars in galaxies of a given mass, we 
first estimate the distribution of galaxy masses by taking the  $K$-band luminosity function 
of \citet{lovedayj00} ($K$-band most closely reflects stellar mass since it is less 
sensitive than other bands to differences in stellar populations) and weighting it by luminosity.
We convert this to stellar mass assuming a constant stellar-mass-to-light 
ratio of $0.5$ \citep{bell01}. This function, plotted in Fig.~\ref{fig:galaxy}, shows 
the amount of stellar mass contributed by galaxies of a given mass --- or assuming 
identical stellar populations --- the fraction of stars 
residing in galaxies of a given stellar mass. 

We estimate the $K$-band luminosity of the Milky Way by converting 
the published $V$-band magnitude of \citet{courteau99} to the $K$-band 
assuming the mean color of an Sbc spiral galaxy from the 2\,MASS Large 
Galaxy Atlas \citep{jarrett03} and applying the color 
conversion from \citep{driver94}. We then convert this to stellar 
mass using the same stellar-mass-to-light ratio used above, i.e., $0.5$. 
In this way we estimate the stellar-mass content of the Milky Way to be
 $10^{10.55 \pm 0.16} = 3.6^{+1.5}_{-1.1} \times 10^{10} M_{\odot}$ 
\citep[see also][]{flynn06}.
Comparing this to the stellar masses of other galaxies
(Fig.~\ref{fig:galaxy}), we find that $77^{+11}_{-14}\%$ of stars reside in 
galaxies less massive than the Milky Way.

%%%%%%%%%%  gal: mass-histogram
%%%%%%%%%%  end fig: group histogram 

\subsection{Host Group Mass}
\begin{figure}[!htb]
  \begin{center}
    \includegraphics[scale=0.55]{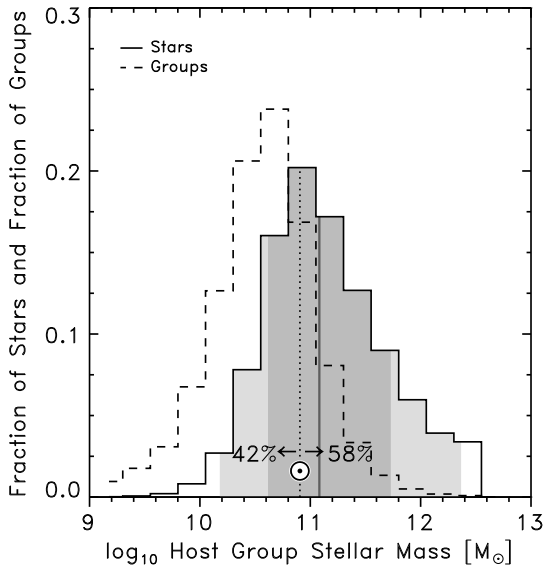}
    \caption{
      The dashed histogram shows the luminosity function of galactic groups (number of groups per 
      interval of B-band luminosity).  The solid histogram shows the luminosity-weighted group 
      luminosity function (approximately the fraction of stars which inhabit a group of given stellar mass).  
      The horizontal axis has been converted to stellar mass assuming a constant B-band stellar-mass-to-light ratio
      of $1.5$ \citep{bell01}. The ``$\odot$'' shows the estimated mass of the Local Group \citep{courteau99} 
      and lies just below the median (vertical grey line).       
    }
    \label{fig:group}
  \end{center}
\end{figure}

The mass of a star's host galactic group or galactic cluster may be correlated with parameters that 
have an influence on habitability. For example, group mass is correlated with the density of the 
galactic environment (number of galaxies per Mpc$^{3}$) that could, like galactocentric radius, be 
associated with the dangers of high stellar densities: ``The presence of a giant elliptical at 
a distance of 50 kpc would have disrupted the Milky Way Galaxy, so that human
beings (and hence astronomers) probably would not have come into existence.'' \citep{vandenberg00}.
Our Local Group of galaxies seems rather typical \citep{vandenberg00} but we would like to quantify this.
Proceeding similarly to our analysis of galaxy mass in Sect. \ref{sec:galaxymass}, we ask,
what fraction of stars live in galactic groups less massive than our Local Group?
Figure \ref{fig:group} shows the luminosity-weighted (i.e.\, stellar mass-weighted) number density of 
galactic groups. The number distribution and luminosity distribution of galactic groups is taken from 
the Two-degree Field Galaxy Redshift Survey Percolation-Inferred Galaxy Group (2PIGG) catalogue 
\citep{Ekev04}. It spans the range from weak groups to rich galaxy clusters.

We estimated the stellar masses of the 2PIGG groups and Local Group galaxies
\citep{courteau99} by converting from the $B$-band assuming a constant
stellar-mass-to-light ratio of 1.5 \citep{bell01}. This gives an estimated stellar mass
of the local group of $10^{10.91 \pm 0.07} = 8.1^{+1.4}_{-1.2} \times 10^{10} M_{\odot}$.
Figure \ref{fig:group} indicates that our Local Group is a typical galactic grouping for a star 
to be part of. Approximately $ 58 \pm 5\%$ of stars live in galactic groups more massive than 
our Local Group. With respect to the mass of its galaxy and the mass of its galactic group, the 
Sun is a fairly typical star in the Universe.

%%%%%%%%%%  group mass: histogram
%%%%%%%%%%  end fig: group density 

\section{Joint Analysis of 11 Solar Properties} \label{sec:jointa}

\subsection{Solar $\chi^{2}$-Analysis}\label{sec:chidf}
We would like to know whether the solar properties, taken as a group, are consistent with noise, 
i.e., are they consistent with  the values of a star selected at random from our 
stellar distributions. We take a $\chi^{2}$ approach to answering this question.
First we estimate the solar $\chi^{2}_{\odot}$, by adding in quadrature, 
for all 11 properties, the differences between the solar values and the median 
stellar values. We find:
\begin{equation}\label{eq:chi2}
  \chi^{2}_{\odot} = \sum_{i=1}^{N=11} \frac{(x_{\odot,i} - \mu_{1/2,i})^2}{\sigma_{68,i}^{2}}  = 7.88^{+0.08}_{-0.30} \\
\end{equation}
where $i$ is the property index, $N=11$ is the number of properties we are considering, 
$\mu_{1/2,i}$ is the median of the $i^{th}$ stellar distribution and $\sigma_{68,i}$ is 
the difference between the median and the upper or lower $68\%$ zone, depending 
on whether the solar value $x_{\odot,i}$ is above or below the median. The uncertainty on 
$\chi^{2}_{\odot}$ is obtained using the uncertainties of $x_{\odot,i}$.

Equation (\ref{eq:chi2}) can be improved on by taking into account {\it i)} the non-Gaussian 
shapes of the stellar distributions and {\it ii)} the larger uncertainties of the medians of 
smaller samples  (our smallest sample is $\sim 100$ stars).  

We employ a bootstrap  analysis \citep{Efron79} to randomly resample data (with replacement) and 
derive a more accurate estimate of $\chi ^2_{\odot}$. Because the bootstrap is a non-parametric 
method, the distributions need not be Gaussian.

\begin{figure}[!t]
  \begin{center}
    \includegraphics[scale=0.55]{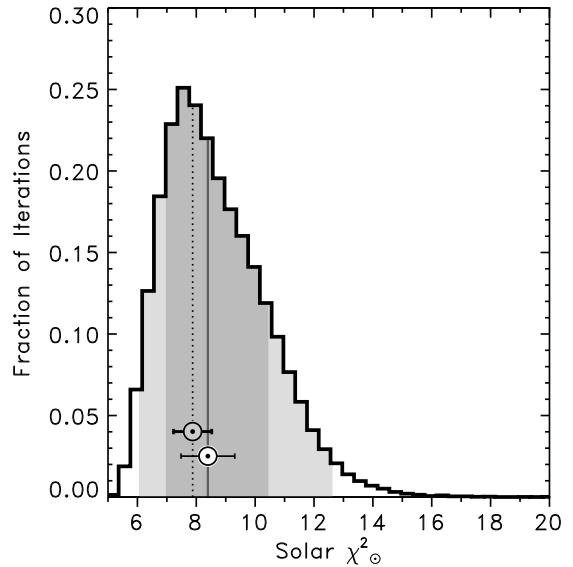}
    \caption{Bootstrapped  solar $\chi^2$ distribution. The median of the distribution 
      (white ``$\odot$'') is $\chi ^2_{\odot} = 8.39 \pm 0.96$. This should be compared to the solar  
      $\chi^2_{\odot}$ value from Eq. \ref{eq:chi2}: $7.88^{+0.08}_{-0.30}$ which is 
      over-plotted (grey ``$\odot$'' on dotted line).}
    \label{fig:sunchi}
  \end{center}
\end{figure}

We obtain $\chi^{2}_{\odot} = 8.39 \pm 0.96$. Figure \ref{fig:sunchi} shows the resulting  
solar $\chi^2$ distribution. The median of this distribution is our adopted  solar $\chi^2$
value. Dividing our adopted solar $\chi^2$ by the number of degrees of freedom gives our
adopted reduced solar $\chi^2$ value:
\begin{equation} \label{eq:redchi}
\chi^{2}_{\odot}/11 = 0.76 \pm 0.09
\end{equation}

The standard conversion of this into a probability of finding a lower $\chi^2$ value 
(assuming normally distributed independent variables) yields:
\begin{equation}   \label{eq:Psimplebetterchi}
P( < \chi^2_{\odot}= 8.39 | N=11) = 0.32 \pm 0.09. 
\end{equation}

%%%%%%%%%%  fig solar sunchi 
%%%%%%%%%%  end fig solar sunchi 

\subsection{Estimate of $P(<\chi^{2}_{\odot})$}\label{sec:starchi}

To quantify how typical the Sun is with respect to our 11 properties, we compare the solar 
$\chi^{2}_{\odot} (=8.39)$ to the distribution of $\chi^{2}$ values obtained from the 
other stars in the samples. 

\begin{figure}[!ht]
        \begin{center}
        \includegraphics[scale=0.55]{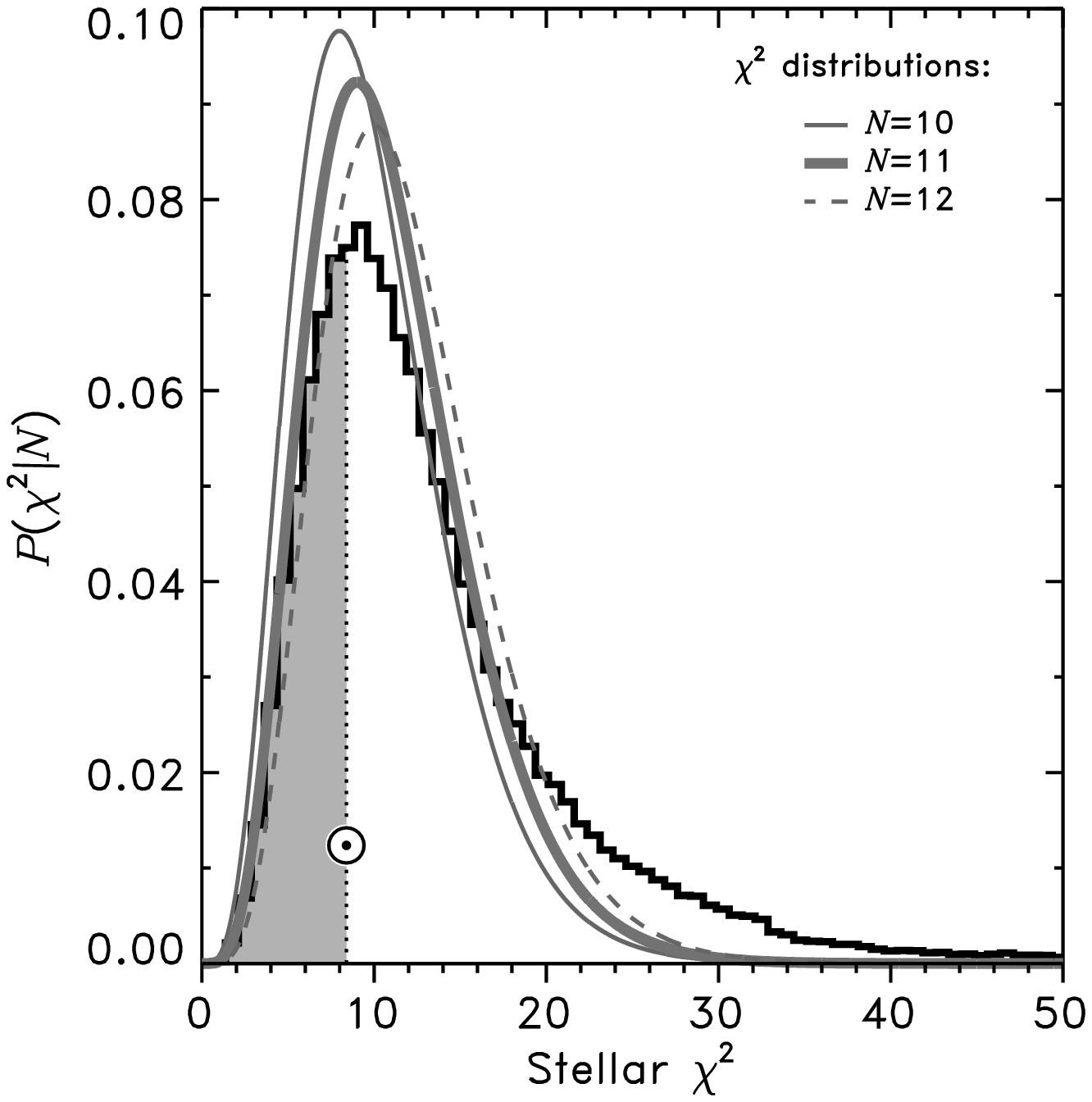}
        \caption{Stellar $\chi^2$ distribution from our Monte Carlo simulation. $P_{\mr{MC}}( < \chi^2_{\odot}= 8.39) = 0.29 \pm 0.11$ 
        (represented by the grey shade) is calculated integrating from $\chi^{2}=0$ to $\chi^{2} = \chi^2_{\odot}$.  
        For comparison, three $\chi^2$ distribution-curves are over-plotted with 10, 11, and 12 degrees
        of freedom.
        The standard probability from the $N=11$ curve yields: $P(< \chi^2_{\odot}= 8.39|N=11) = 0.32 \pm 0.09$.
      }
                \label{fig:chiprob}
        \end{center}
\end{figure}

We perform a Monte Carlo simulation \citep{Metropolis49} to calculate an estimate 
of each star's $\chi^2$ value ($\chi^2_{\star}$). 
The histogram shown in Figure \ref{fig:chiprob} is the resulting Monte Carlo stellar $\chi^2$ distribution. 
Three standard $\chi^2$ distributions have been over-plotted for comparison ($N=10,11,12$).
The probability of finding a star with $\chi^2$ lower than or equal to solar is:
\begin{equation} \label{eq:mcprob}
P_{\mr{MC}}( \leq \chi^2_{\odot}=8.39 | N=11) = 0.29 \pm 0.11
\end{equation}
The Monte Carlo $\chi^2$  distribution has a similar shape to the standard 
$\chi^2$ distribution function for $N=11$, and thus both yield similar 
probabilities: $P_{\mr{MC}}(\leq \chi^{2}) = 0.29  \sim P( \leq \chi^{2}) = 0.32$ 
(Eqs.~\ref{eq:Psimplebetterchi} and \ref{eq:mcprob}). The more appropriate Monte Carlo
distribution has a longer tail, produced by the longer super-Gaussian tails of the stellar distributions.

Table \ref{table:chiimprovedp} summarizes our analysis for the solar $\chi^{2}_{\odot}$ values and the
probabilities $P(<\chi^{2}_{\odot}$). 
Our simple $\chi^{2}_{\odot} = 7.88$ estimate increased to $8.39$ 
and the uncertainty increased by a factor of $\sim 3$
after non-Gaussian and sample size effects were included as additional sources of uncertainty.
Our improved analysis yields $P_{\mr{MC}}(\leq\chi_{\odot}^{2})$, with a longer tail and brings the probability down
from $0.32 \pm 0.09$ to $0.29 \pm 0.11$. If this value were close to 1, almost all other stars would have lower $\chi^2$ values and we
would have good reason to suspect that the Sun is not a typical star.
However, this preliminary low value of $0.29$  indicates that if a star is chosen at random,  
the probability that it will be more typical ($\sim$ have a lower $\chi^{2}$ value) 
than the Sun (with respect to the 11 properties analysed here), is only  $29 \pm 11 \%$.
The details of our improved estimates of $\chi^{2}_{\odot}$ and $P(<\chi^{2}_{\odot})$
can be found in the Appendix \ref{sec:appxprob}.

%%%%%%%%%%  fig stellar chi 
%%%%%%%%%%  end fig stellar chi 

%%%%%%%%%%%%%  table with chi and P(<chi) values for simple and improved estiamtes   %%%%%%%%%%%%%%%% 
\begin{deluxetable*}{ccccc}
\tablewidth{0pt}
\tablecaption{Summary of $\chi^{2}$ and $P(< \chi^{2}_{\odot})$ results.}
\tablehead{
\colhead{Analysis}&\colhead{$\chi^{2}_{\odot}$}&\colhead{$\chi^{2}_{\odot}/11$}&\colhead{$P(<\chi^{2}_{\odot}| N=11$)}&\colhead{$P_{\mr{MC}}(<\chi^{2}_{\odot}| N=11)$}  
}
\startdata 
simple    &   $7.88^{+0.08}_{-0.30}$ (Eq.~\ref{eq:chi2})       &  $0.72^{+0.01}_{-0.03}$   &   $0.28^{+0.01}_{-0.03}$ (Eq.~\ref{eq:Psimple})  &  ---\\
improved  &   $8.39 \pm 0.96$              &  $0.76 \pm 0.09$ (Eq.~\ref{eq:redchi})         &   $0.32  \pm 0.09$ (Eq.~\ref{eq:Psimplebetterchi})       &  $0.29 \pm 0.11$ (Eq.~\ref{eq:mcprob})  \\
% \\ \tableline
\enddata
\label{table:chiimprovedp} 
\end{deluxetable*}
%%%%%%%%%%%%%  end table
%%%%%%%%%%  sun2stars  Result figures
\begin{figure*}[!b]
  \begin{center}
    \begin{tabular}{rr}                           
                \includegraphics[scale=0.55]{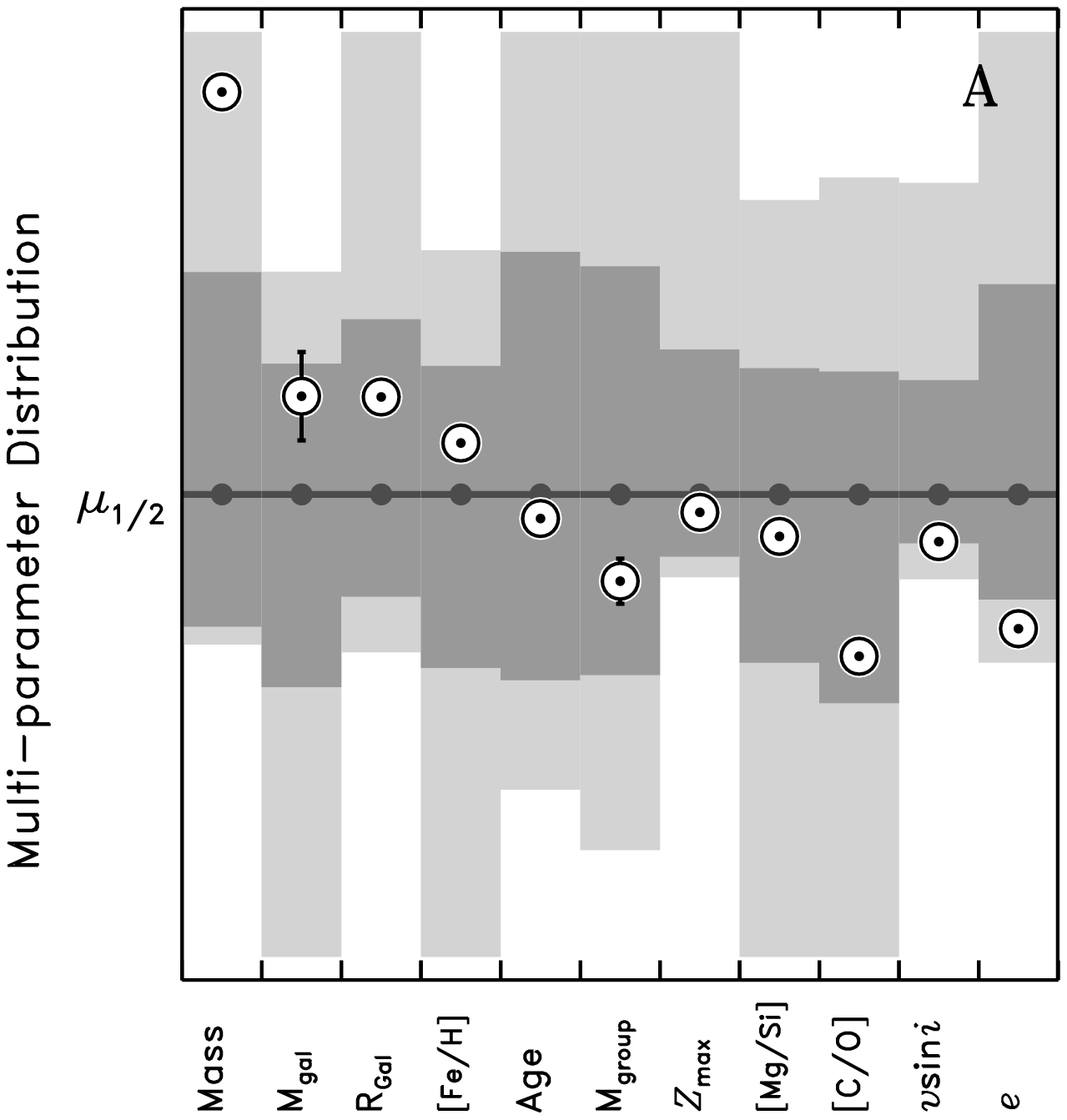} &
                \includegraphics[scale=0.55]{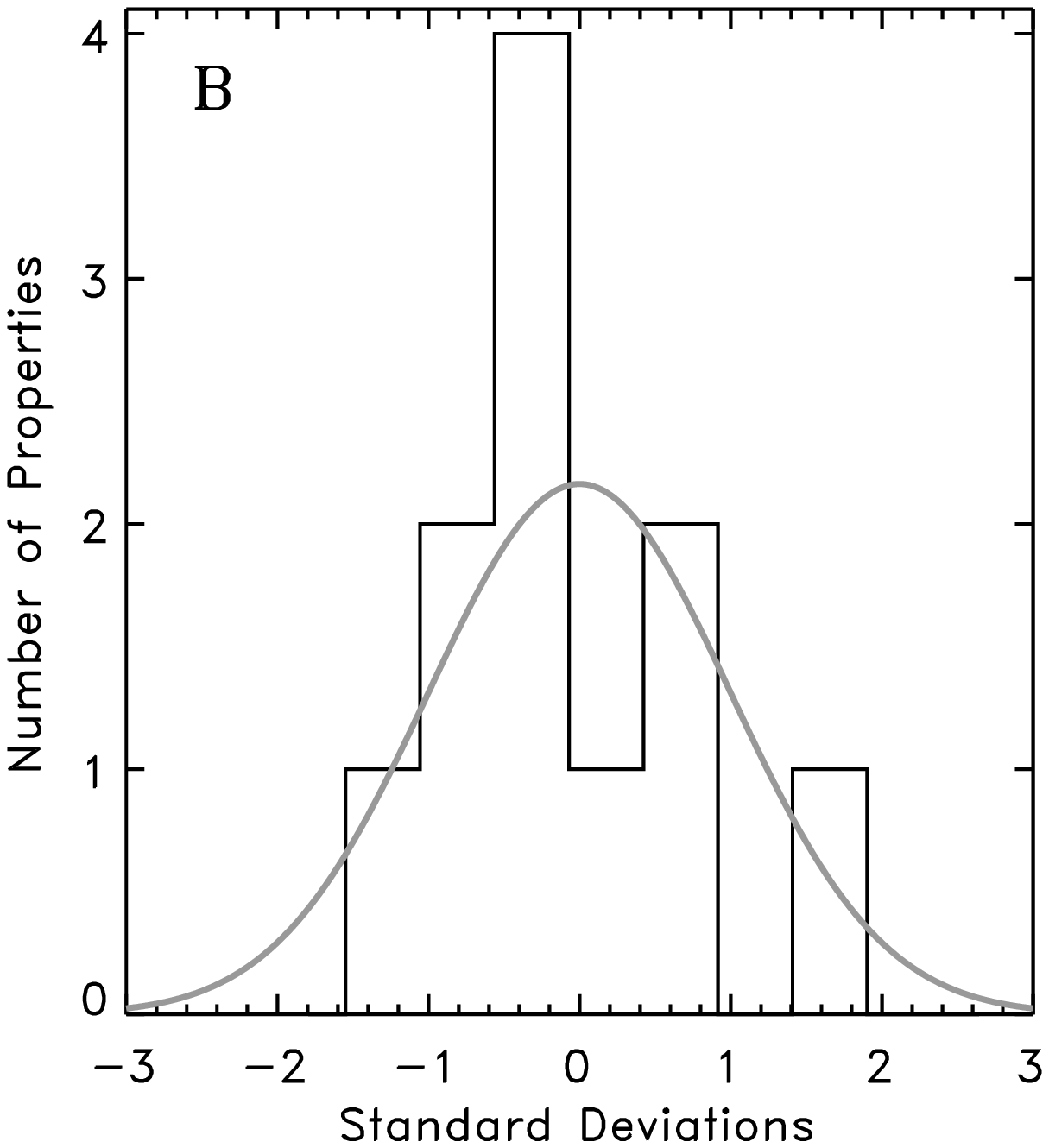} \\
                \includegraphics[scale=0.55]{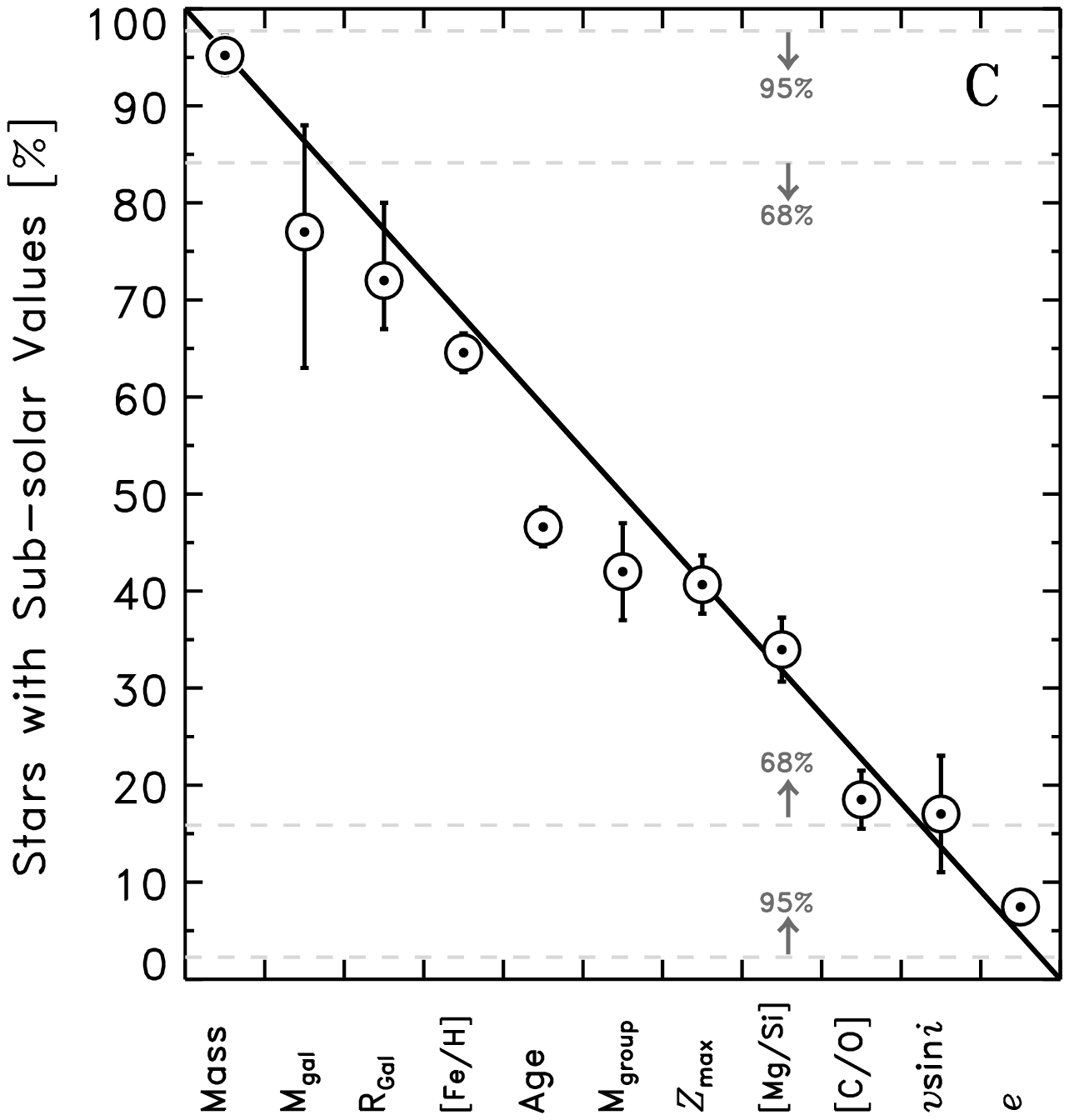} &
                \includegraphics[scale=0.55]{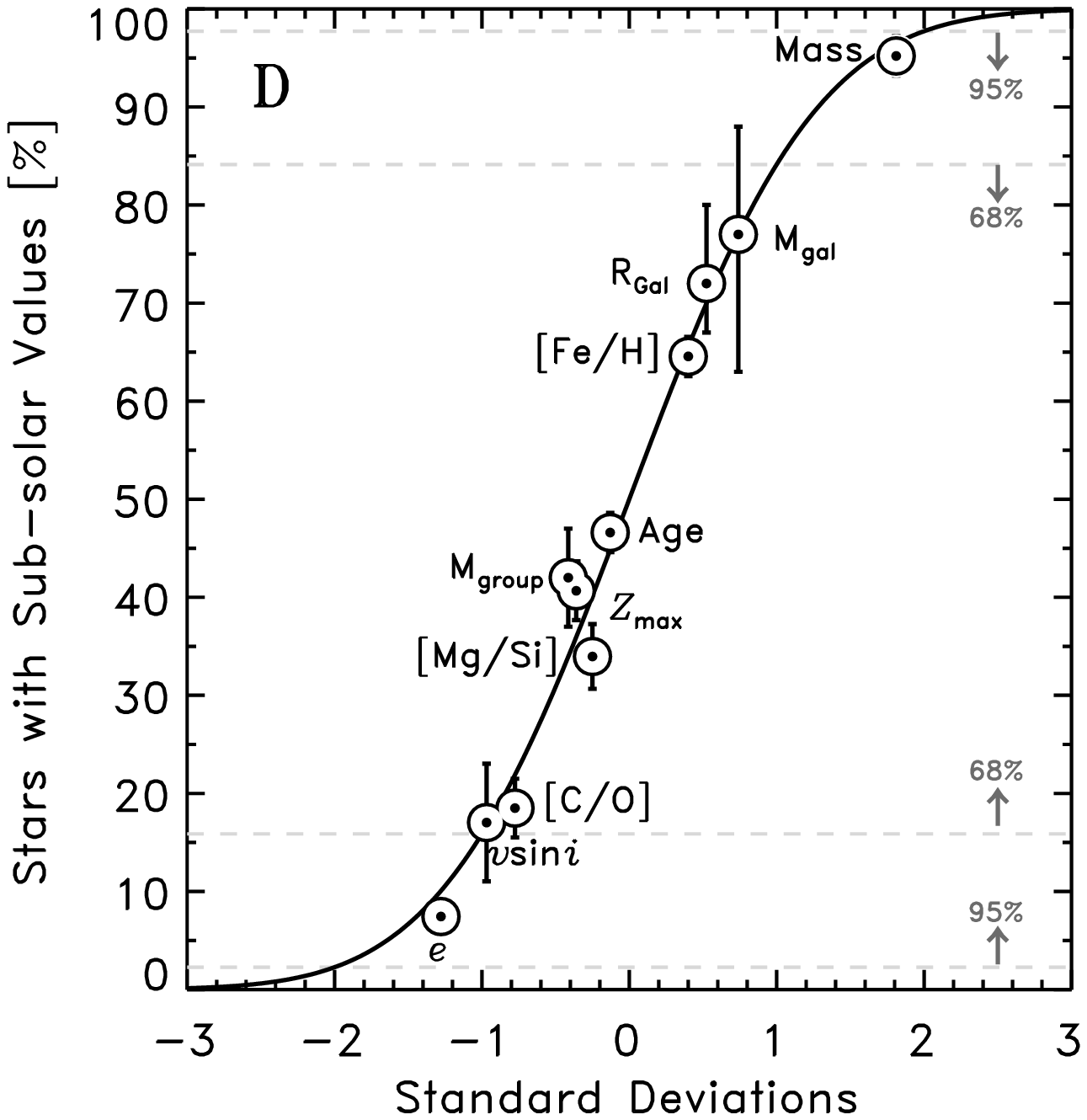}  \\
        \end{tabular}
        \caption{Various representations of our main results.   
        \textbf{A}: Solar values of 11 properties compared to the distribution for each property
         Each distribution's median value is indicated by a small filled circle.  The dark and light 
         grey shades represent the 68\% and 95\% zones respectively.
         \textbf{B}: Histogram of the number of properties as a function of the number of standard 
         deviations the solar value is from the median of that property. The grey curve is a 
         Gaussian probability distribution normalised to 11 parameters.
        \textbf{C}: Percentage $n_{i}\%$ of stars with sub-solar values as a function of property.
         The average signal expected from a random star is shown by the solid line 
         (see Sec.~\ref{sec:results}).
         \textbf{D}: Percentage $n_{i}\%$ of stars with sub-solar values as a function of 
         the number of standard deviations the solar value is from the median of that property.
         The solid curve is a cumulative Gaussian distribution --- if every sample were a 
         Gaussian distribution, every solar dot would sit exactly on the line. 
         Just as in (\textbf{C}), the dashed lines encompass the 68\% and 95\% zones.
         Similar to the results from  Figure \ref{fig:chiprob}, these four panels indicate that the 
         Sun is a typical star.}
                \label{fig:sum}
        \end{center}
\end{figure*}
%%%%%%%%%%  end fig: sunstars 

\section{Results} \label{sec:results}

Figure \ref{fig:sum} shows four different representations of our results. 
Figure \ref{fig:sum}{\bf A} compares the solar values to each stellar distribution's median and 68\% and 95\% zones. 
The Sun lies beyond the 68\% zone for three properties: mass ($95\%$), eccentricity ($93\%$) and 
rotational velocity ($88\%$).
No solar property lies beyond the 95\% zone.
The histogram in Figure\ref{fig:sum}{\bf B} is the distribution of solar values in units of standard deviations:
\begin{equation}\label{eq:stdevs}
       z_i = \frac{x_{\odot,i}-\mu_{1/2,i}}{\sigma_{68,i}}
\end{equation}
For each stellar property $i$, the Sun has a larger value than $n_i\%$ of the stars.
If the Sun were a randomly selected
star, we would expect the percentages $n_i\%$ to be scattered roughly evenly between 0\% and 100\%.
\clearpage
When the $n_i\%$ values are lined up in decreasing order (Fig.~\ref{fig:sum}{\bf C}), we expect them to 
be near the line given by:
\begin{equation}\label{eq:noiseline}
        n_{i,\mr{expected}}\% = \Big[1-\frac{(i-1/2)}{N}\Big] \times 100\%
\end{equation}
and plotted in Figure\ref{fig:sum}{\bf C}.
Any anomalies would show up as points `$\odot$' significantly distant from the line.

Figure\ref{fig:sum}{\bf D}) compares the percentages $n_i\%$ of stars
having sub-solar values (shown in Figure\ref{fig:sum}{\bf C}) with the solar values expressed in units of
standard deviations from each 
distribution's median (shown in Fig.~\ref{fig:sum}{\bf B}).  If the stellar distributions were 
perfect Gaussians, the translation from $z_i$ to
$n_i$ would be given by the cumulative Gaussian distribution (Fig.~\ref{fig:sum}{\bf D}, \textit{solid curve}).
That the points lie along this line demonstrates that the approximation of our
distributions as Gaussians is reasonable. 

Table \ref{table:sum} lists percentages $n_i\%$ of stars for each property
(as shown in Fig.~\ref{fig:sum}). In the lower half of the table we list
properties not included in this analysis because of correlations with properties that are included.

Individual stellar uncertainties make the observed characteristic widths ($\sigma_{68}$, Table \ref{table:samples1}) larger than
the widths of the intrinsic distributions. This broadening effect makes the Sun appear more
typical than it really is when $\sigma_{68}$ and the individual stellar uncertainties ($\sigma_{\star}$) are 
of similar size and the individual stellar uncertainties are much larger than the solar
uncertainty ($\sigma_{\odot}$). We estimate that our results are not significantly affected 
by this broadening effect.

Our resulting probability of finding a star with a $\chi^2$ lower or equal to the solar value of 
$29 \pm 11\%$ (Eq.~\ref{eq:mcprob}), is consistent with the probability we would obtain if stellar 
multiplicity were included in our study. Using the volume limited sample used for stellar mass in 
Section~\ref{sec:mass} (125 A1--M7 stars within 7.1 pc) the probability that a randomly selected star 
will be single is $52.8 \pm 4.5\%$, which means that $\sim$ half of stars are single while the other 
half have one or more companions. Including this in our bootstrap analysis and Monte Carlo simulations 
(see Appendix~\ref{sec:appmult}) marginally increases the probability in Eq.~\ref{eq:mcprob} to 
$33 \pm 11\%$. If the multiplicity data for 246 G dwarfs from \cite{duquennoy91} is used instead --- the 
probability that a randomly selected G dwarf  will be single is $37.8 \pm 2.9 \%$ --- then the probability 
in Eq.~\ref{eq:mcprob} would increase to $34 \pm 11\%$. The inclusion of stellar multiplicity marginally
increases our reported probability.

In Figures 6 and 7 of \cite{radick98}, the Sun's short-term variability as a function
of average chromospheric activity, appears $\sim 1 \sigma$ low, compared to a distribution 
of 35 F3--K7 Sun-like stars \citep{lockwood97}. \cite{lockwood07} suggest that the
Sun's small total irradiance variation compared to stars with similar mean chromospheric activity,
may be due to their limited sample and the lack of solar observations out of the Sun's equatorial plane.
We do not include short or long term variability (chromospheric or photometric) in Table \ref{table:sum}
because of the small size of the \cite{lockwood07} sample. We also do not
include the chromospheric index $R'_{\mr{HK}}$ (see Table \ref{table:sum}, bottom panel)
as one of our 11 properties because of its correlation with the chromospheric ages of our sample.

%%%%%%%%%%%%%  table with summary vals %%%%%%%%%%%%%%%% 
\begin{deluxetable*}{lcll}
  \tabletypesize{\footnotesize}
  \tablewidth{0pt}
  \tablecaption{Summary of How the Sun Compares to Other Stars (see Fig.~\ref{fig:sum})} 
  \tablehead{
    \colhead{Parameter}           & \colhead{Figure}         &  \colhead{$n_i\%$}    &  \colhead{Level of Anomaly}  
  }
  \startdata 
  Mass               &\ref{fig:mass}      & $95 \pm 2\%$       &  of nearby stars are less massive than the Sun. \\
  Age                &\ref{fig:age}       & $53 \pm 2\%$       &  of stars in the thin disk of the Galaxy are older than the Sun. \\ 
  $[$Fe/H$]$         &\ref{fig:fe}        & $65 \pm 2\%$       &  of nearby stars are more iron-poor than the Sun. \\
  $[$C/O$]$          &\ref{fig:ratios}A   & $81 \pm 3\%$       &  of nearby stars have a higher C/O ratio than the Sun. \\
  $[$Mg/Si$]$        &\ref{fig:ratios}B   & $66 \pm 3\%$       &  of nearby stars have a higher Mg/Si ratio than the Sun. \\
  $v\sin i$          &\ref{fig:vsini}     & $83 \pm 7\%$       &  of nearby Sun-like-mass stars rotate faster than the Sun. \\
  $e$                &\ref{fig:e}         & $93 \pm 1\%$       &  of nearby stars have larger galactic orbital eccentricities than the Sun. \\
  $Z_{\mr{max}}$     &\ref{fig:z}         & $59 \pm 3\%$       &  of nearby stars reach farther from the Galactic plane than the Sun. \\
  R$_{\mr{Gal}}$     &\ref{fig:radius}    & $72^{+8}_{-5}\%$   &  of stars in the Galaxy are closer to the galactic center than the Sun. \\
  M$_{\mr{gal}}$     &\ref{fig:galaxy}    & $77^{+11}_{-14}\%$ &  of stars in the Universe are in galaxies less massive than the Milky Way. \\
  M$_{\mr{group}}$   &\ref{fig:group}     & $58 \pm 5\%$       &  of stars in the Universe are in groups more massive than the local group. \\
  \tableline
  \cutinhead{Properties Not Included in the Analysis Because They Are Correlated with the Selected 11 Parameters}   
  Mass: IMF$_{\mr{Stellar}}$ &\ref{fig:mass}       &  $94 \pm 2\%$    &  of nearby stars are less massive than the Sun. \\
  Age: SFR$_{\mr{Cosmic}}$   &\ref{fig:age}        &  $86 \pm 5\%$    &  of stars in the Universe are older than the Sun. \\ 
  Age\tablenotemark{a}                    &  ---                &  $55 \pm 2\%$    &  of nearby Sun-like-mass stars are older than the Sun. \\ 
  $[$Fe/H$]$\tablenotemark{b}               &  ---                &  $56 \pm 5\%$    &  of nearby stars are more iron-poor than the Sun. \\
  $v\sin i$\tablenotemark{c}                &  ---                &  $92 \pm 5\%$    &  of nearby Sun-like-mass stars rotate faster than the Sun. \\
  $\log R'_{\mr{HK}}$\tablenotemark{d}        &  ---                &  $51 \pm 2\%$    &  of nearby FGKM stars are more chromospherically active. \\
  $[$O/Fe$]$                 &  ---                &  $75\pm 3\%$     &  of nearby stars have a lower O/Fe ratio than the Sun. \\
  $R_{\mr{min}}$             &  ---                &  $91\pm 1\%$     &  of nearby stars get closer to the Galactic center. \\
  $v_{\mr{LSR}}$             &  ---                &  $93 \pm 1\%$    &  of nearby stars have smaller velocity with respect to the LSR. \\
  $|U|$                      &  ---                &  $75 \pm 1\%$    &  of nearby stars have larger absolute radial velocity. \\
  $|V|$                      &  ---                &  $82 \pm 1\%$    &  of nearby stars have larger absolute tangential velocity. \\
  $|W|$                      &  ---                &  $58 \pm 1\%$    &  of nearby stars have larger absolute vertical velocity. \\
  \enddata
    \tablenotetext{a}{1126 stars (A5--K2)  from \cite{Nordstrom04}.}
    \tablenotetext{b}{91 stars (GK) from \cite{Favata97}.}
    \tablenotetext{c}{590 stars (F8--K2) from \cite{Nordstrom04}.}
    \tablenotetext{d}{866 stars (FGKM) from \cite{Wright04}.}
\label{table:sum} 
\end{deluxetable*}
%%%%%%%%%%%%%%%%%%%%%%%%%%%%%%%%%%%%%%%%%%%%%%%%%%%%%%%% 

\newpage
\section{Discussion and Interpretation} \label{sec:discuss}

The probability $P_{\mr{MC}}(\leq \chi^2_{\odot}) = 0.29 \pm 0.11$ classifies the Sun as a typical star.
How robust is this result? The probability of finding a star with a $\chi^2$ lower than or equal 
to $\chi^2_{\odot}$, 
depends on the properties selected for the analysis (see problem {\it iii} of Section I). 
For example, if we had chosen to consider only 
mass and eccentricity data, this analysis would yield  
$P_{\mr{MC}}(\chi^2 \le \chi^2_{\odot}) = 0.94 \pm 0.4$, 
i.e.,\ the Sun would appear mildly ($\sim 2 \sigma$) anomalous. 
If on the other hand, we had chosen to remove mass and eccentricity from the
analysis, we would obtain $P_{\mr{MC}}(\chi^2 \le \chi^2_{\odot}) = 0.07 \pm 0.04$, 
which is anomalously low.
The most common cause of such a result is the over-estimation of 
error bars.  The next most common cause is the preselection of properties known to have
$n_{i}\% \sim 50\%$.

\cite{Gustafsson98} discussed the atypically large solar mass, and proposed
an anthropic explanation --- the Sun's high mass 
is probably related to our own existence.
He suggested that the solar mass could hardly have been greater than $\sim 1.3 \, \mr{M}_{\odot}$
since the main sequence lifetime of a $1.3 \, \mr{M}_{\odot}$ star is $\sim  5$ billion 
years \citep{clayton83}. He also discussed how the dependence of the width of the 
circumstellar habitable zone on the host star's mass probably favours host stars 
within the mass range $0.8$--$1.3\, \mr{M}_{\odot}$.

Our property selection criteria is to have the largest number of maximally independent properties 
that have a plausible correlation with habitability and, ones for which a representative stellar sample 
could be assembled.
Our joint analysis does not weight any parameter more heavily than any other.
If the only properties associated with habitability are mass and eccentricity 
then we have diluted a $\sim 2 \sigma$ signal that would be consistent with Gustafsson's proposed 
anthropic explanation.

Our analysis points in another direction.
If mass and eccentricity were the only properties associated with habitability, then the 
solar values for the remaining nine
properties would be consistent with noise. However, a joint analysis
of just the remaining nine properties produces a $\chi^{2}_{\odot,9} = 3.6 \pm 0.4$ and
the anomalously low probability: $P(\le \chi^{2}_{\odot,9}) = 0.07 \pm 0.04$, which suggests that 
the nine properties are unlikely to be the properties of a star selected at random
with respect to these properties.

The $\chi^{2}$ fit of the 11 points  in Panel {\bf C} of Fig.~13 to the diagonal
line yields a fit that is
substantially better then the fit of the remaining nine properties to Eq. 7 with $N=9$.
In other words, the joint analyis suggests that although mass and eccentricity are  
the most anomalous solar properties, it is unlikely that they are associated with habitability, because
without them, it is unlikely that the remaining solar properties are just noise.
Thus,
the Sun, despite its mildly ($\sim 2 \sigma$) anomalous mass and eccentricity, can be 
considered a typical, randomly selected star.

There may be stellar properties crucial for life that were not tested here.
If we have left out the most important properties, with respect to which the Sun is 
atypical, then our Sun-is-typical conclusion will not be valid.
If we have sampled all properties associated with habitability, 
our Sun-is-typical result suggests that there are no special requirements on a star for it
to be able to 
host a planet with life.
\newpage

%%%%%%%%%%%%%%%%%%%%%%%%%%%%%%%%%%%%%%%%%%%%%%%
\section{Conclusions}\label{sec:conclusions}
We have compared the Sun to representative stellar samples for 11 properties. 
Our main results are as follows:
\begin{itemize}{}
\item Stellar mass and Galactic orbital eccentricity are the most anomalous properties. 
The Sun is more massive than $95 \pm2\%$ of nearby stars and has a 
Galactic orbital eccentricity lower than $93\pm1\%$ FGK stars within 40 pc. 
\item Our joint bootstrap analysis yields a solar $\chi^2$ $\chi^{2}_{\odot} = 8.39 \pm 0.96$ and
a solar reduced $\chi^2$ $\chi^{2}_{\odot}/11=  0.76 \pm 0.09$.  The probability of 
finding a star with a $\chi^2$ lower than or equal to solar
$P_{\mr{MC}}(  \leq \chi^2_{\odot}=8.39\pm 0.96) = 0.29 \pm 0.11$. 
\end{itemize}
To our knowledge, this is the most comprehensive and quantitative 
comparison of the Sun with other stars.
We find that taking all 11 properties together, the Sun is a typical star. 
This finding is largely in agreement with \citet{Gustafsson98}, however
our results undermine the proposition that an anthropic explanation is needed
for the comparatively large mass of the Sun.

Further work could encompass the inclusion of other properties potentially 
associated with habitability. Another improvement would come when 
larger stellar samples become available for which
all properties could be derived, instead of using different samples for different properties as
was done here.
In addition, research in the molecular evolution that led to the origin of life may, in the future, 
be able to provide more clues as to which stellar properties might be
associated with our existence on Earth, orbiting the Sun.

\acknowledgments
Acknowledgments: We would like to thank Charles Jenkins for clarifying discussions of statistics,
particularly on how to include stellar multiplicity, and Martin Asplund and Jorge Mel\'endez for discussions 
of elemental abundances. JAR acknowledges an RSAA PhD research scholarship. MP acknowledges the financial 
support of the Australian Research Council. EG acknowledges the financial support of the Finnish Cultural Foundation.
%%%%%%%%%%%%%%%%%%%%%%%%%%%%%%%%%%%%%%%%%%%%%%%
\section*{Appendix A}
% \appendix
\section*{PROPERTY CORRELATIONS}\label{sec:appendix}

The $\chi^{2}$ formalism and the use of the
$\chi^{2}$-distribution to obtain $P(< \chi^{2}_{\odot}|N)$, --- improved using Monte-Carlo simulations
in Section \ref{sec:starchi} to obtain $P_{\mr{MC}}(\leq \chi^2_{\odot})$ --- assumes that each parameter 
is independent of the others. In selecting our 11 properties we have selected properties which are maximally
independent based on plotting
property 1 vs property 2 for the same stars.
We show seven such plots in this Appendix.

If there are correlations between the analysed properties, then the 
number of degrees of freedom $N$ could drop from 11 to $\sim 10.5$  (see Fig.~12).
Some properties have been excluded from the analysis due to a correlation with 
another property in the analysis. 

\section*{A1. ELEMENTAL RATIOS}

\begin{figure}[!ttt]
     \begin{center}
        \includegraphics[scale=0.55]{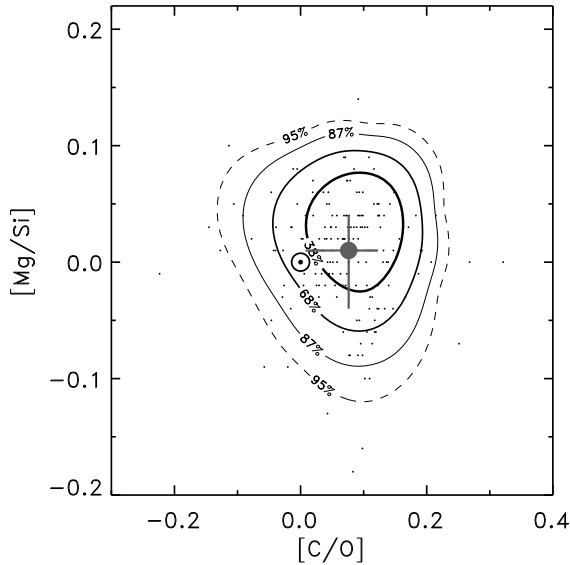} 
                \caption{Carbon to oxygen ratio [C/O] versus magnesium to silicon ratio [Mg/Si]
                  of 176 FG stars with abundances for these elements \citep{Reddy03}. In 
                  Figs.~\ref{fig:ratios}C and \ref{fig:ratios}D we showed that the [C/O] and [Mg/Si] distributions are largely independent 
                  of [Fe/H]. Here we show that these distributions are also largely independent of each other.
                  Note that in this comparison we only use the data from \cite{Reddy03}, since it is the largest
                 available sample with C, O, Mg and Si abundances.
                }
                \label{fig:comgsi}
        \end{center}
\end{figure}

In Figure \ref{fig:comgsi} we show the distribution for carbon to oxygen ratio
[C/O] versus the magnesium to silicon ratio [Mg/Si] of 176 FG stars.

%%%%%%%%%%  fig: co vs mg
%%%%%%%%%%  end fig: co vs mg

\section*{A2. MASS, AGE, AND ROTATIONAL VELOCITY}
\begin{figure*}[!tttt]
  \begin{center}
    \begin{tabular}{rr}
      \includegraphics[scale=0.55]{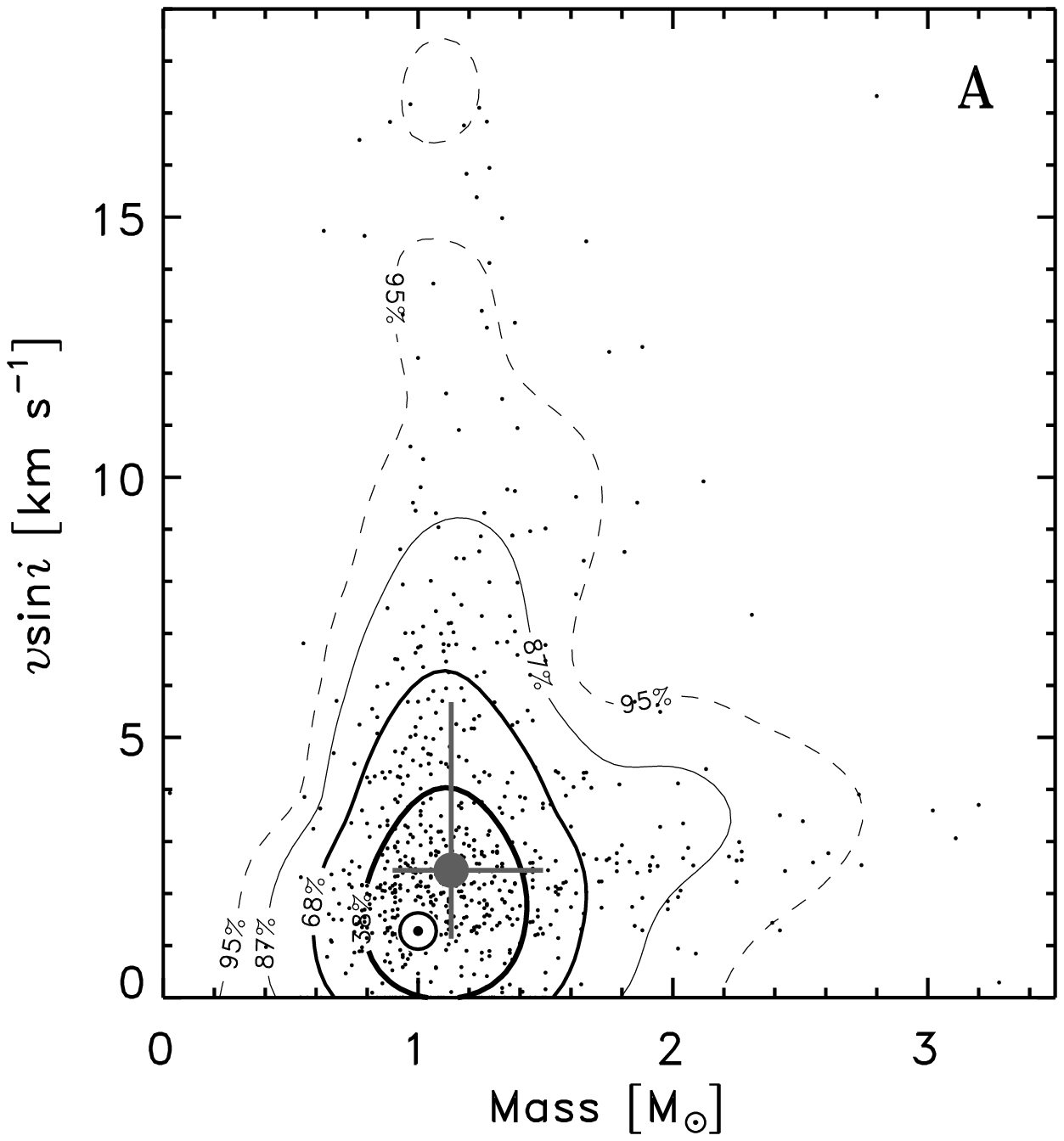} & \includegraphics[scale=0.55]{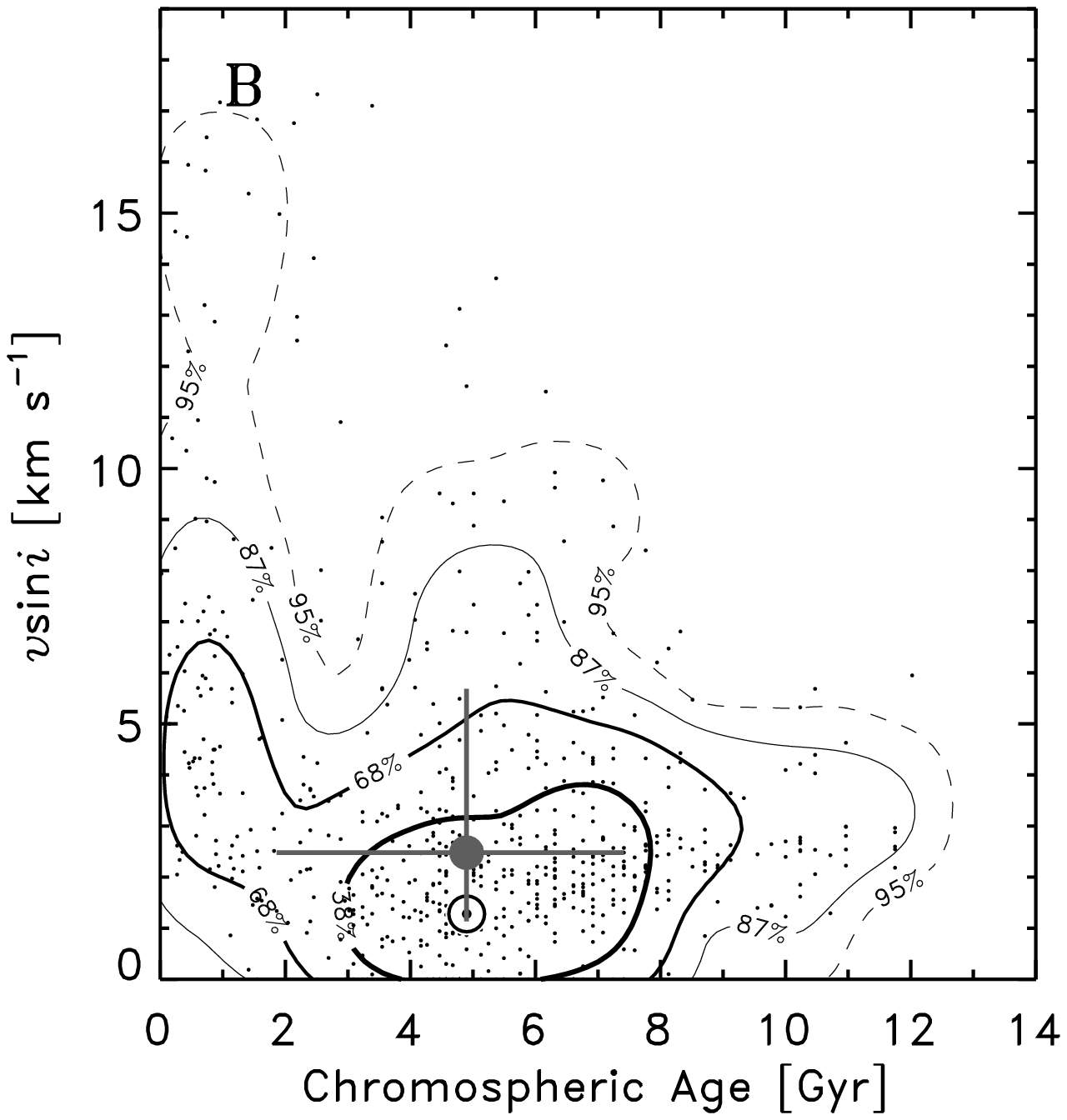} \\
      \includegraphics[scale=0.55]{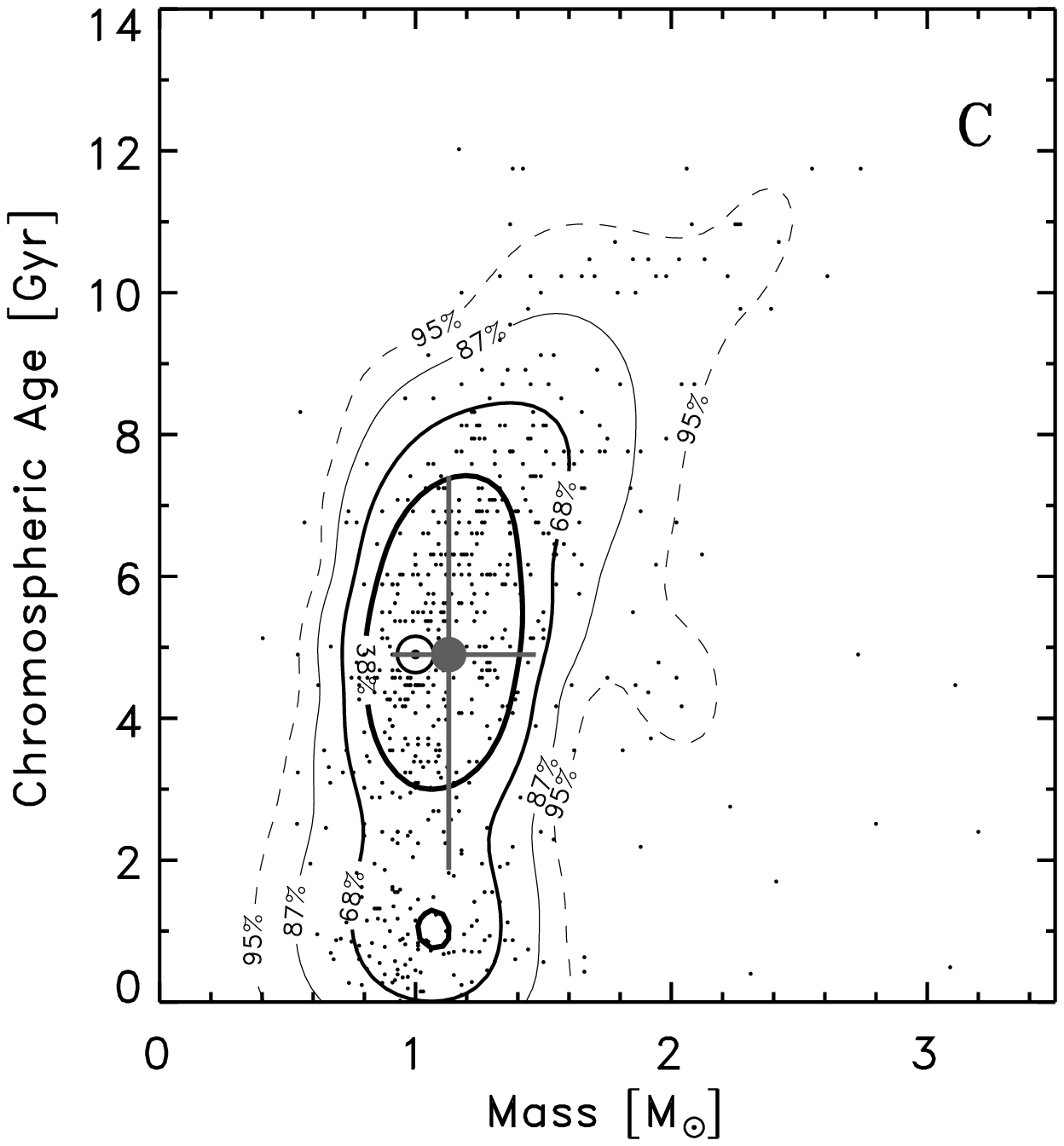} & \includegraphics[scale=0.55]{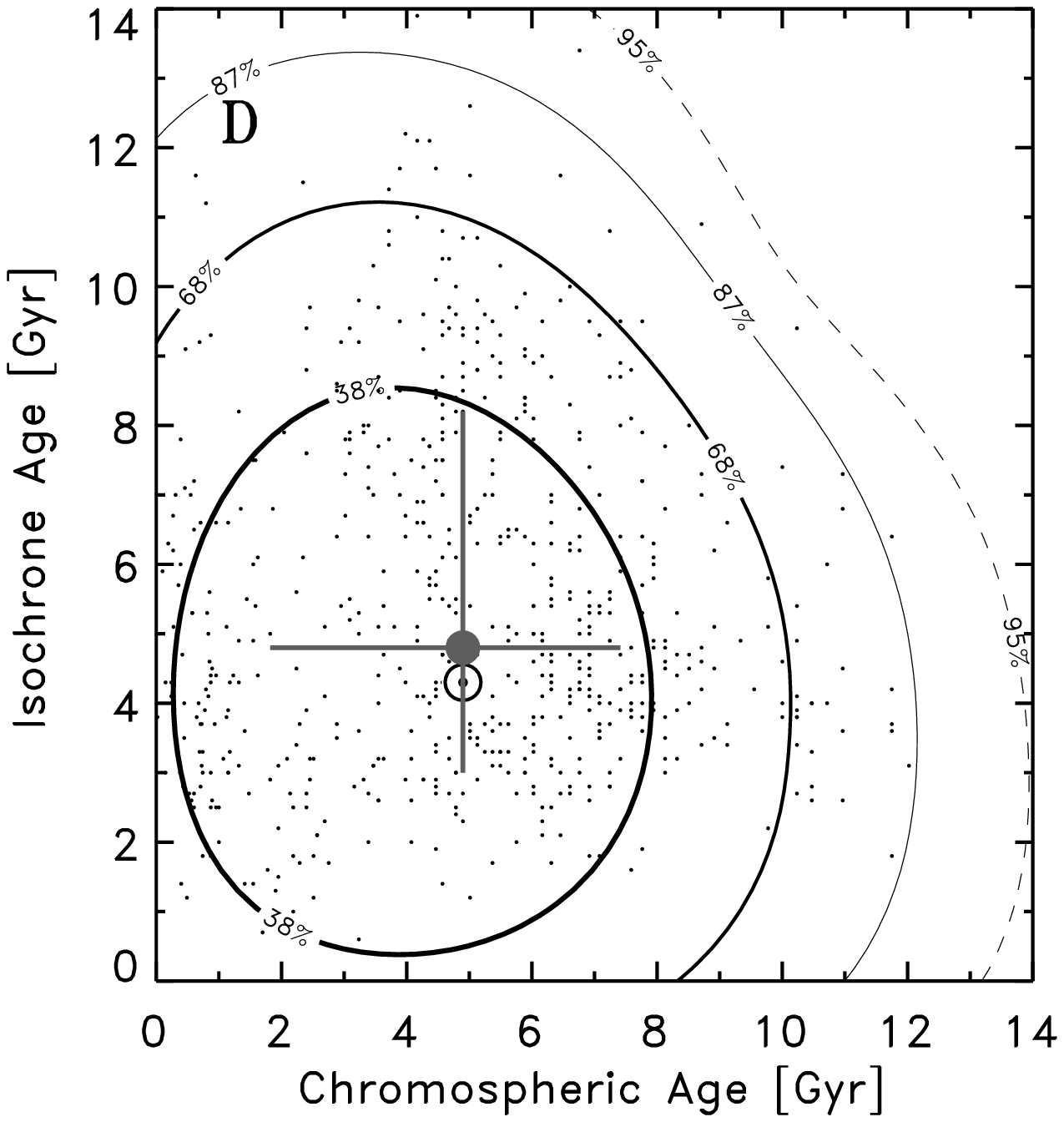} \\
    \end{tabular}
    \caption{Correlation plots between various properties. 
      For all four panels we use the stars common to both \cite{Wright04} and \cite{Valenti05}.
      Panel ({\bf A}): mass vs rotational velocity $v \sin i$ for 713 FGK stars. 
      This panel shows the degree of correlation between mass and $v \sin i$.  See \cite{gray05} for
a stronger correlation between these two variables when a larger mass range and more active stars 
are kept in the sample.
      To minimize the effect of this correlation on our analysis, we restrict the range of mass in Fig.~5 to 
$0.9$ to $1.1 M_{\odot}$.
      Panel ({\bf B}): chromospheric age versus  $v\sin i$ for 641 FGK stars. 
The lack of 
      correlation between chromospheric determined ages and rotational velocities is shown.
      Panel ({\bf C}): no strong correlation between mass and chromospheric age for 639 FGK stars. 
      Panel ({\bf D}): the ages of 637 stars determined
      by the chromospheric method  versus their ages from the isochrone method.
    }
    \label{fig:massagev}
  \end{center}
\end{figure*}

In Figure \ref{fig:massagev} we show four correlation plots for mass, chromospheric 
age, rotational velocity and $v \sin i$. We use the stars common to both \cite{Wright04} 
and \cite{Valenti05} for which these observables are available.

%%%%%%%%%%  fig: mass, age vs vsini
%%%%%%%%%%  end fig:mass age vs vsini

\section*{A3. GALACTIC ORBITAL PARAMETERS}

\begin{figure*}[!tttt]
     \begin{center}
       \begin{tabular}{cc}
         \includegraphics[scale=0.55]{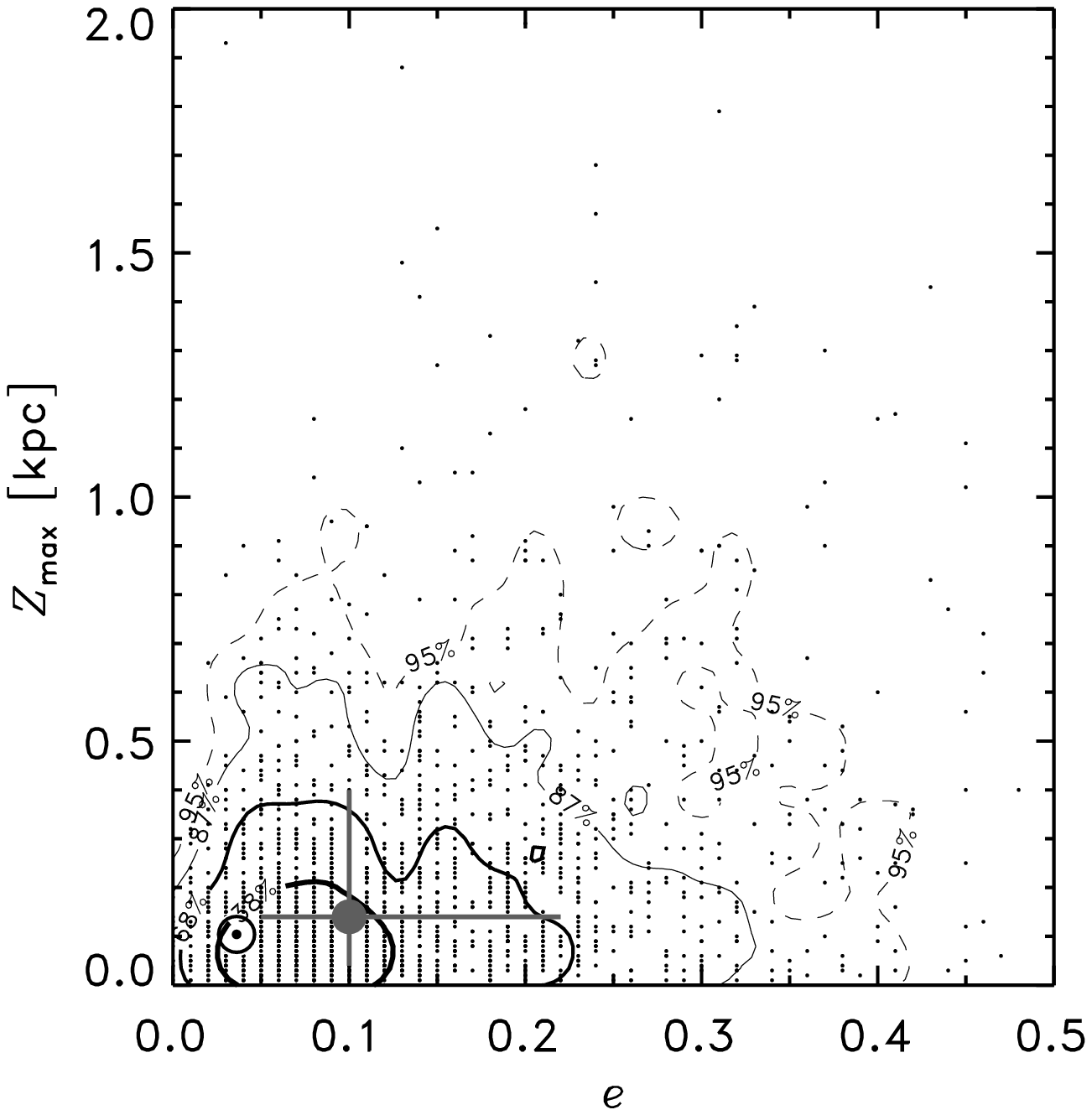} & \includegraphics[scale=0.55]{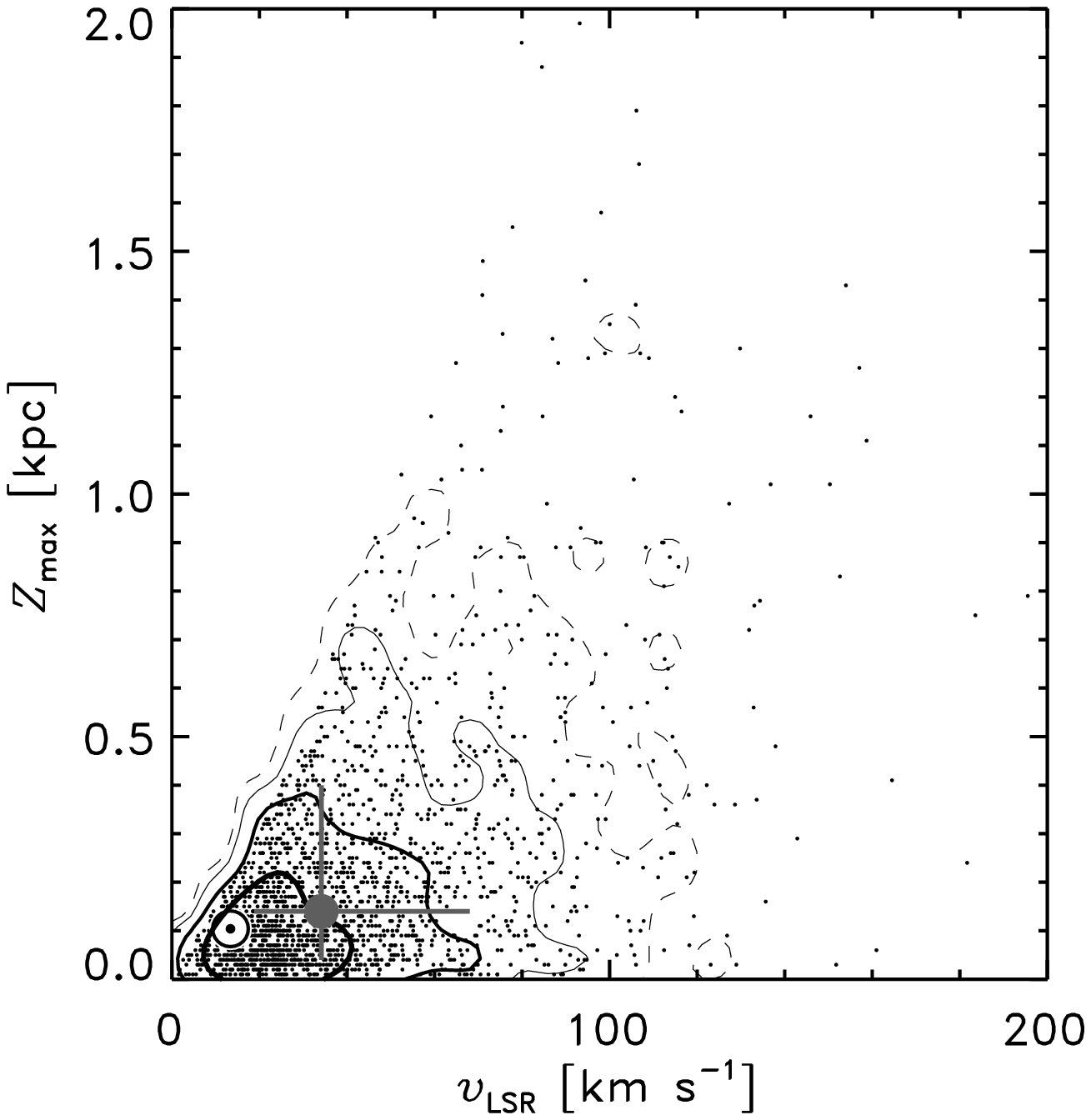}  
       \end{tabular}
                \caption{
                  Left panel: Galactic orbital eccentricity $e$ versus $Z_{\mr{max}}$ for 1987 FGK 
                  stars within 40 pc \citep{Nordstrom04}. 
                  The orbital eccentricity is not correlated with $Z_{\mr{max}}$.  
                  Right panel: $v_{\mr{LSR}}$ versus $Z_{\mr{max}}$ for the same stars. Because $v_{\mr{LSR}}$
                  is more strongly correlated with $Z_{\mr{max}}$ than eccentricity, eccentricity has been
                  selected for the joint analysis instead of $v_{\mr{LSR}}$.
                  As in Fig.~\ref{fig:ratios}, the contours correspond to 38\%, 68\%, 82\% and 95\%.
                }
                \label{fig:ezmax}
        \end{center}
\end{figure*}

The Galactic orbital eccentricity ($e$) and the magnitude of the 
galactic orbital velocities with respect to the local standard of rest ($v_{\mr{LSR}}$) are
strongly correlated (see Fig.~\ref{fig:e} in Sec.~\ref{sec:orbital}).
We selected $e$ instead of $v_{\mr{LSR}}$ because of its near independence of the 
maximum height above the galactic plane ($Z_{\mr{max}}$); see Figure \ref{fig:ezmax}.

%%%%%%%%%%  fig: e vs Zm
%%%%%%%%  end fig: e vs Zm

\section*{APPENDIX B}

\section*{IMPROVED ESTIMATES OF  $\chi^{2}_{\odot}$ AND  $P(<\chi^{2}_{\odot})$}\label{sec:appxprob}

In Section \ref{sec:starchi}, with 11 degrees of freedom, the reduced $\chi^2$ from
Equation \ref{eq:mcprob} is $\chi^{2}_{\odot} \; / \; 11 = 0.72^{+0.01}_{-0.03}$. 
Since $\chi^{2}_{\odot} \; / \; 11 < 1$, the Sun's properties are consistent with
the Sun being a randomly selected star.

To improve on this preliminary analysis (but with a similar conclusion), as mentioned in Section \ref{sec:starchi},
we employ a bootstrap  analysis \citep{Efron79} to randomly resample data (with replacement) and 
derive a more accurate estimate of $\chi ^2_{\odot}$. Because the bootstrap is a non-parametric 
method, the distributions need not be Gaussian.

For every iteration, each parameter's stellar distribution is randomly resampled and a $\chi ^2_{\odot}$ value 
is calculated using Eq.~(\ref{eq:chi2}). The uncertainties $\sigma_{\odot,i}$ of the solar values $x_{\odot,i}$ 
are also included in the bootstrap method: for every iteration, the Solar value for each parameter
is replaced in Eq.~(\ref{eq:chi2}) by a randomly selected value from a normal distribution with median 
$\mu_{1/2,i}=x_{\odot,i}$ and standard deviation $\sigma_{\odot,i}$.
The process was iterated 100,000 times, although the resulting distribution varies very little once 
the number of iterations reaches $\sim10,000$.

The median of this distribution and the error on the median yields our improved value for the 
reduced $\chi ^2_{\odot}$ (Fig.~\ref{fig:sunchi}). The uncertainty of the median of each re-sampled 
distribution varies inversely proportionally to the square root of the number of stars in the 
distribution, $\Delta \mu_{1/2,i} \propto 1/\sqrt{N_{\star,i}}.$ In other words, median values 
are less certain for smaller samples and this uncertainty is included in our improved estimate 
of $\chi^{2}_{\odot}$,  and its uncertainty.

We find the probability of finding a star with a $\chi^2_{\star}$ value lower than 
the solar $\chi^{2}_{\odot}$, for $N=11$ degrees of freedom in the standard way \citep{press92} 
and obtain: 
\begin{equation}   \label{eq:Psimple}
P( < \chi^2_{\odot}=7.88^{+0.08}_{-0.30}|11) = 0.28^{+0.01}_{-0.03}
\end{equation}
To improve our estimate of the probability of finding a star with lower $\chi^2$ value than the Sun,
we perform a  Monte Carlo simulation \citep{Metropolis49} to calculate an estimate 
of each star's $\chi^2$ value ($\chi^2_{\star}$). For every iteration, we randomly select 
a star from each stellar distribution. We then calculate its $\chi^2_{\star}$ value by replacing 
the solar value $x_{\odot,i}$ with that star's value $x_{\star,i}$ in Eq.~(\ref{eq:chi2}). 
This process was repeated 100,000 times to create our Monte Carloed stellar $\chi^2$ distribution.
Stars were randomly selected with replacement, thus the simulated $\chi^2$ distribution accounts for 
small number statistics and non-Gaussian distributions. The probability of finding a star with 
$\chi^2$ lower than or equal to solar is $P_{\mr{MC}}=0.29 \pm 0.11$.

The results of our analysis for the Solar $\chi^{2}_{\odot}$ values and the
probabilities $P(<\chi^{2}_{\odot}$) are summarized in Table \ref{table:chiimprovedp}

\section*{B1. ADDITION OF A DISCRETE PARAMETER} \label{sec:appmult}
In Section~\ref{sec:results} we discuss the addition of stellar multiplicity to our analysis.  
Since stellar multiplicity cannot easily be approximated by a one-sided Gaussian (particularly because 
the Sun is on the edge of the distribution, i.e., it is of multiplicity one), we modified our 
Monte Carlo procedure to include this discrete parameter. The likelihood of observing  a 
particular $\chi^2$ for the 11 parameters is
\begin{equation}
\exp\left(-\frac{1}{2}\sum_{i=1}^{11} \chi^2_i \right).
\end{equation}

We take the probability $p(1)$ of a star being a single star, to be  $53.8 \pm 4.5 \%$, obtained 
from our sample of nearby stars (Sec.~\ref{sec:mass}). The likelihood $L$ of observing  a particular $\chi^2$ 
and $p(1)$ is the  product
\begin{equation}
L = p(1) \exp \left(-\frac{1}{2} \sum_{i=1}^{11} \chi^2_i \right).
\end{equation}
Taking logarithms we can then compute the distribution of the statistic $S$, where
\begin{equation}
S = \ln p(1)  -\frac{1}{2} \sum_{i= 1}^{11} \chi^2_i.
\end{equation}
The distribution of $S$ allows us to obtain the results for multiplicity reported at the end of Section~\ref{sec:results}.

%%%%%%%%%%%%%%%%%%%%%%%%%%%%%%%%%%%%%%%%%%%%%%%
%BIBLIOGRAPHY  AND  END  DOCUMENT
% Using bibtex
\bibliographystyle{apj}
% \bibliography{references}

\begin{thebibliography}{68}
\expandafter\ifx\csname natexlab\endcsname\relax\def\natexlab#1{#1}\fi

\bibitem[{{All{\`e}gre} {et~al.}(1995){All{\`e}gre}, {Manh{\`e}s}, \&
  {G{\"o}pel}}]{Allegre95}
{All{\`e}gre}, C.~J., {Manh{\`e}s}, G., \& {G{\"o}pel}, C. 1995, \gca, 59, 1445

\bibitem[{{Allende Prieto}(2006)}]{Allendeprieto06}
{Allende Prieto}, C. 2006, ArXiv Astrophysics e-prints

\bibitem[{{Allende Prieto} {et~al.}(2001){Allende Prieto}, {Lambert}, \&
  {Asplund}}]{allendeprieto01}
{Allende Prieto}, C., {Lambert}, D.~L., \& {Asplund}, M. 2001, \apjl, 556, L63

\bibitem[{{Asplund} {et~al.}(2005){Asplund}, {Grevesse}, \&
  {Sauval}}]{Asplund05}
{Asplund}, M., {Grevesse}, N., \& {Sauval}, A.~J. 2005, in ASP Conf. Ser. 336:
  Cosmic Abundances as Records of Stellar Evolution and Nucleosynthesis, 25

\bibitem[{{Bahcall} \& {Soneira}(1980)}]{Bahcall80}
{Bahcall}, J.~N., \& {Soneira}, R.~M. 1980, \apjs, 44, 73

\bibitem[{{Bell} \& {de Jong}(2001)}]{bell01}
{Bell}, E.~F., \& {de Jong}, R.~S. 2001, \apj, 550, 212

\bibitem[{{Bensby} \& {Feltzing}(2006)}]{Bensby06}
{Bensby}, T., \& {Feltzing}, S. 2006, \mnras, 367, 1181

\bibitem[{{Bensby} {et~al.}(2005){Bensby}, {Feltzing}, {Lundstr{\"o}m}, \&
  {Ilyin}}]{Bensby05}
{Bensby}, T., {Feltzing}, S., {Lundstr{\"o}m}, I., \& {Ilyin}, I. 2005, \aap,
  433, 185

\bibitem[{{Bertelli} \& {Nasi}(2001)}]{bertelli01}
{Bertelli}, G., \& {Nasi}, E. 2001, \aj, 121, 1013

\bibitem[{{Carter}(1983)}]{Carter83}
{Carter}, B. 1983, Philos. Trans.R. Soc. London, A, 310

\bibitem[{{Clayton}(1983)}]{clayton83}
{Clayton}, D.~D. 1983, {Principles of stellar evolution and nucleosynthesis}
  (Chicago: University of Chicago Press, 1983)

\bibitem[{{Courteau} \& {van den Bergh}(1999)}]{courteau99}
{Courteau}, S., \& {van den Bergh}, S. 1999, \aj, 118, 337

\bibitem[{{Dehnen} \& {Binney}(1998)}]{dehnen98}
{Dehnen}, W., \& {Binney}, J.~J. 1998, \mnras, 298, 387

\bibitem[{{Driver} {et~al.}(1994){Driver}, {Phillipps}, {Davies}, {Morgan}, \&
  {Disney}}]{driver94}
{Driver}, S.~P., {Phillipps}, S., {Davies}, J.~I., {Morgan}, I., \& {Disney},
  M.~J. 1994, \mnras, 268, 393

\bibitem[{{Duquennoy} \& {Mayor}(1991)}]{duquennoy91}
{Duquennoy}, A., \& {Mayor}, M. 1991, \aap, 248, 485

\bibitem[{{Edvardsson} {et~al.}(1993{\natexlab{a}}){Edvardsson}, {Andersen},
  {Gustafsson}, {Lambert}, {Nissen}, \& {Tomkin}}]{Edvardsson93}
{Edvardsson}, B., {Andersen}, J., {Gustafsson}, B., {Lambert}, D.~L., {Nissen},
  P.~E., \& {Tomkin}, J. 1993{\natexlab{a}}, \aap, 275, 101

\bibitem[{{Edvardsson} {et~al.}(1993{\natexlab{b}}){Edvardsson}, {Andersen},
  {Gustafsson}, {Lambert}, {Nissen}, \& {Tomkin}}]{Edvardsson93b}
---. 1993{\natexlab{b}}, \aaps, 102, 603

\bibitem[{{Efron}(1979)}]{Efron79}
{Efron}, B. 1979, The Annals of Statistics, 7, 1

\bibitem[{{Eisenhauer} {et~al.}(2005){Eisenhauer}, {Genzel}, {Alexander},
  {Abuter}, {Paumard}, {Ott}, {Gilbert}, {Gillessen}, {Horrobin}, {Trippe},
  {Bonnet}, {Dumas}, {Hubin}, {Kaufer}, {Kissler-Patig}, {Monnet},
  {Str{\"o}bele}, {Szeifert}, {Eckart}, {Sch{\"o}del}, \&
  {Zucker}}]{Eisenhauer05}
{Eisenhauer}, F. {et~al.} 2005, \apj, 628, 246

\bibitem[{{Eke} {et~al.}(2004){Eke}, {Frenk}, {Baugh}, {Cole}, {Norberg},
  {Peacock}, {Baldry}, {Bland-Hawthorn}, {Bridges}, {Cannon}, {Colless},
  {Collins}, {Couch}, {Dalton}, {de Propris}, {Driver}, {Efstathiou}, {Ellis},
  {Glazebrook}, {Jackson}, {Lahav}, {Lewis}, {Lumsden}, {Maddox}, {Madgwick},
  {Peterson}, {Sutherland}, \& {Taylor}}]{Ekev04}
{Eke}, V.~R. {et~al.} 2004, \mnras, 355, 769

\bibitem[{{Favata} {et~al.}(1996){Favata}, {Micela}, \& {Sciortino}}]{Favata96}
{Favata}, F., {Micela}, G., \& {Sciortino}, S. 1996, \aap, 311, 951

\bibitem[{{Favata} {et~al.}(1997){Favata}, {Micela}, \& {Sciortino}}]{Favata97}
---. 1997, \aap, 323, 809

\bibitem[{{Feltzing} {et~al.}(2001){Feltzing}, {Holmberg}, \&
  {Hurley}}]{Feltzing01}
{Feltzing}, S., {Holmberg}, J., \& {Hurley}, J.~R. 2001, \aap, 377, 911

\bibitem[{{Flynn} {et~al.}(2006){Flynn}, {Holmberg}, {Portinari}, {Fuchs}, \&
  {Jahrei{\ss}}}]{flynn06}
{Flynn}, C., {Holmberg}, J., {Portinari}, L., {Fuchs}, B., \& {Jahrei{\ss}}, H.
  2006, \mnras, 372, 1149

\bibitem[{{Flynn} {et~al.}(1996){Flynn}, {Sommer-Larsen}, \&
  {Christensen}}]{flynn96}
{Flynn}, C., {Sommer-Larsen}, J., \& {Christensen}, P.~R. 1996, \mnras, 281,
  1027

\bibitem[{{Fuhrmann}(2008)}]{fuhrmann08}
{Fuhrmann}, K. 2008, \mnras, 384, 173

\bibitem[{{Gonzalez}(1999{\natexlab{a}})}]{Gonzalez99a}
{Gonzalez}, G. 1999{\natexlab{a}}, \mnras, 308, 447

\bibitem[{{Gonzalez}(1999{\natexlab{b}})}]{Gonzalez99b}
---. 1999{\natexlab{b}}, Astronomy and Geophysics, 40, 25

\bibitem[{{Gonzalez} {et~al.}(2001){Gonzalez}, {Brownlee}, \&
  {Ward}}]{Gonzalez01}
{Gonzalez}, G., {Brownlee}, D., \& {Ward}, P. 2001, Icarus, 152, 185

\bibitem[{{Gould} {et~al.}(1996){Gould}, {Bahcall}, \& {Flynn}}]{Gould96}
{Gould}, A., {Bahcall}, J.~N., \& {Flynn}, C. 1996, \apj, 465, 759

\bibitem[{{Gray}(2005)}]{gray05}
{Gray}, D.~F. 2005, {The Observation and Analysis of Stellar Photospheres} (The
  Observation and Analysis of Stellar Photospheres, 3rd Edition, by D.F.~Gray.~
  ISBN 0521851866.~http://www.cambridge.org/us/ \\
  /catalogue/catalogue.asp?isbn=0521851866.~Cambridge, UK: Cambridge University
  Press, 2005.)

\bibitem[{{Grether} \& {Lineweaver}(2006)}]{Grether06}
{Grether}, D., \& {Lineweaver}, C.~H. 2006, \apj, 640, 1051

\bibitem[{{Grether} \& {Lineweaver}(2007)}]{Grether07}
---. 2007, \apj, 669, 1220

\bibitem[{{Gustafsson}(1998)}]{Gustafsson98}
{Gustafsson}, B. 1998, Space Science Reviews, 85, 419

\bibitem[{{Gustafsson} {et~al.}(1999){Gustafsson}, {Karlsson}, {Olsson},
  {Edvardsson}, \& {Ryde}}]{Gustafsson99}
{Gustafsson}, B., {Karlsson}, T., {Olsson}, E., {Edvardsson}, B., \& {Ryde}, N.
  1999, \aap, 342, 426

\bibitem[{{Henry}(2006)}]{Henry06}
{Henry}, T.~J. 2006, RECONS database

\bibitem[{{Hernandez} {et~al.}(2000){Hernandez}, {Valls-Gabaud}, \&
  {Gilmore}}]{hernandez00}
{Hernandez}, X., {Valls-Gabaud}, D., \& {Gilmore}, G. 2000, \mnras, 316, 605

\bibitem[{{Hogg} {et~al.}(2005){Hogg}, {Blanton}, {Roweis}, \&
  {Johnston}}]{hogg05}
{Hogg}, D.~W., {Blanton}, M.~R., {Roweis}, S.~T., \& {Johnston}, K.~V. 2005,
  \apj, 629, 268

\bibitem[{{Hopkins} \& {Beacom}(2006)}]{Hopkins06}
{Hopkins}, A.~M., \& {Beacom}, J.~F. 2006, \apj, 651, 142

\bibitem[{{Jarrett} {et~al.}(2003){Jarrett}, {Chester}, {Cutri}, {Schneider},
  \& {Huchra}}]{jarrett03}
{Jarrett}, T.~H., {Chester}, T., {Cutri}, R., {Schneider}, S.~E., \& {Huchra},
  J.~P. 2003, \aj, 125, 525

\bibitem[{{Kasting} {et~al.}(1993){Kasting}, {Whitmire}, \&
  {Reynolds}}]{kasting93}
{Kasting}, J.~F., {Whitmire}, D.~P., \& {Reynolds}, R.~T. 1993, Icarus, 101,
  108

\bibitem[{{Kroupa}(2002)}]{Kroupa02}
{Kroupa}, P. 2002, Science, 295, 82

\bibitem[{{Kroupa} \& {Weidner}(2005)}]{Kroupa05}
{Kroupa}, P., \& {Weidner}, C. 2005, in ASSL Vol. 327: The Initial Mass
  Function 50 Years Later, ed. E.~{Corbelli}, F.~{Palla}, \& H.~{Zinnecker},
  175

\bibitem[{{Kuchner} \& {Seager}(2005)}]{Kuchner05}
{Kuchner}, M.~J., \& {Seager}, S. 2005, ArXiv Astrophysics e-prints

\bibitem[{{Lineweaver} {et~al.}(2004){Lineweaver}, {Fenner}, \&
  {Gibson}}]{Lineweaver04}
{Lineweaver}, C.~H., {Fenner}, Y., \& {Gibson}, B.~K. 2004, Science, 303, 59

\bibitem[{{Lockwood} {et~al.}(2007){Lockwood}, {Skiff}, {Henry}, {Henry},
  {Radick}, {Baliunas}, {Donahue}, \& {Soon}}]{lockwood07}
{Lockwood}, G.~W., {Skiff}, B.~A., {Henry}, G.~W., {Henry}, S., {Radick},
  R.~R., {Baliunas}, S.~L., {Donahue}, R.~A., \& {Soon}, W. 2007, \apjs, 171,
  260

\bibitem[{{Lockwood} {et~al.}(1997){Lockwood}, {Skiff}, \&
  {Radick}}]{lockwood97}
{Lockwood}, G.~W., {Skiff}, B.~A., \& {Radick}, R.~R. 1997, \apj, 485, 789

\bibitem[{{Loveday}(2000)}]{lovedayj00}
{Loveday}, J. 2000, \mnras, 312, 557

\bibitem[{{Metropolis} \& {Ulam}(1949)}]{Metropolis49}
{Metropolis}, N., \& {Ulam}, S. 1949, Journal of the American Statistical
  Association, 44, 335

\bibitem[{{Metzger} {et~al.}(1998){Metzger}, {Caldwell}, \&
  {Schechter}}]{Metzger98}
{Metzger}, M.~R., {Caldwell}, J.~A.~R., \& {Schechter}, P.~L. 1998, \aj, 115,
  635

\bibitem[{{Meyer}(1975)}]{meyer75}
{Meyer}, S.~L. 1975, {Data Analysis for Scientists and Engineers} (Data
  Analysis for Scientists and Engineers, by Stuart L.~Meyer p.~186.~ISBN
  0471599956.~NY, USA: John Wiley \& Sons, 1975.)

\bibitem[{{Nordstr{\"o}m} {et~al.}(2004){Nordstr{\"o}m}, {Mayor}, {Andersen},
  {Holmberg}, {Pont}, {J{\o}rgensen}, {Olsen}, {Udry}, \&
  {Mowlavi}}]{Nordstrom04}
{Nordstr{\"o}m}, B. {et~al.} 2004, \aap, 418, 989

\bibitem[{{Press} {et~al.}(1992){Press}, {Teukolsky}, {Vetterling}, \&
  {Flannery}}]{press92}
{Press}, W.~H., {Teukolsky}, S.~A., {Vetterling}, W.~T., \& {Flannery}, B.~P.
  1992, {Numerical recipes in FORTRAN. The art of scientific computing}
  (Cambridge: University Press, |c1992, 2nd ed.)

\bibitem[{{Radick} {et~al.}(1998){Radick}, {Lockwood}, {Skiff}, \&
  {Baliunas}}]{radick98}
{Radick}, R.~R., {Lockwood}, G.~W., {Skiff}, B.~A., \& {Baliunas}, S.~L. 1998,
  \apjs, 118, 239

\bibitem[{{Ram{\'{\i}}rez} {et~al.}(2007){Ram{\'{\i}}rez}, {Allende Prieto}, \&
  {Lambert}}]{ramirez07}
{Ram{\'{\i}}rez}, I., {Allende Prieto}, C., \& {Lambert}, D.~L. 2007, \aap,
  465, 271

\bibitem[{{Reddy} {et~al.}(2003){Reddy}, {Tomkin}, {Lambert}, \& {Allende
  Prieto}}]{Reddy03}
{Reddy}, B.~E., {Tomkin}, J., {Lambert}, D.~L., \& {Allende Prieto}, C. 2003,
  \mnras, 340, 304

\bibitem[{{Reid}(2002)}]{Reid02}
{Reid}, I.~N. 2002, \pasp, 114, 306

\bibitem[{{Reid} {et~al.}(2007){Reid}, {Turner}, {Turnbull}, {Mountain}, \&
  {Valenti}}]{reid07}
{Reid}, I.~N., {Turner}, E.~L., {Turnbull}, M.~C., {Mountain}, M., \&
  {Valenti}, J.~A. 2007, \apj, 665, 767

\bibitem[{{Rocha-Pinto} {et~al.}(2000{\natexlab{a}}){Rocha-Pinto}, {Maciel},
  {Scalo}, \& {Flynn}}]{Rochapinto00b}
{Rocha-Pinto}, H.~J., {Maciel}, W.~J., {Scalo}, J., \& {Flynn}, C.
  2000{\natexlab{a}}, \aap, 358, 850

\bibitem[{{Rocha-Pinto} {et~al.}(2000{\natexlab{b}}){Rocha-Pinto}, {Scalo},
  {Maciel}, \& {Flynn}}]{Rochapinto00a}
{Rocha-Pinto}, H.~J., {Scalo}, J., {Maciel}, W.~J., \& {Flynn}, C.
  2000{\natexlab{b}}, \apjl, 531, L115

\bibitem[{{Schechter}(1976)}]{schechter76}
{Schechter}, P. 1976, \apj, 203, 297

\bibitem[{{Silk}(2007)}]{silk07}
{Silk}, J. 2007, Astronomy and Geophysics, 48, 30

\bibitem[{{Soderblom}(1983)}]{soderblom83}
{Soderblom}, D.~R. 1983, \apjs, 53, 1

\bibitem[{{Soderblom}(1985)}]{soderblom85}
---. 1985, \aj, 90, 2103

\bibitem[{{Truran} \& {Heger}(2005)}]{Truran05}
{Truran}, Jr., J.~W., \& {Heger}, A. 2005, {Origin of the Elements}
  (Meteorites, Comets and Planets: Treatise on Geochemistry, Volume 1)

\bibitem[{{Valenti} \& {Fischer}(2005)}]{Valenti05}
{Valenti}, J.~A., \& {Fischer}, D.~A. 2005, \apjs, 159, 141

\bibitem[{{van den Bergh}(2000)}]{vandenberg00}
{van den Bergh}, S. 2000, {The Galaxies of the Local Group} (The galaxies of
  the Local Group, by Sidney Van den Bergh.~Published by Cambridge, UK:
  Cambridge University Press, 2000 Cambridge Astrophysics Series Series, vol
  no: 35, ISBN: 0521651816.)

\bibitem[{{Wright} {et~al.}(2004){Wright}, {Marcy}, {Butler}, \&
  {Vogt}}]{Wright04}
{Wright}, J.~T., {Marcy}, G.~W., {Butler}, R.~P., \& {Vogt}, S.~S. 2004, \apjs,
  152, 261

\end{thebibliography}
% Directly inserted in .tex document

\end{document}